\documentclass[fleqn,usenatbib]{mnras}

\usepackage{newtxtext,newtxmath}

\usepackage[T1]{fontenc}

\DeclareRobustCommand{\VAN}[3]{#2}
\let\VANthebibliography\thebibliography
\def\thebibliography{\DeclareRobustCommand{\VAN}[3]{##3}\VANthebibliography}

\usepackage{graphicx}	
\usepackage{amsmath}	
\usepackage{siunitx}

\newcommand{\referee}[1]{#1}

\title[Baryon inventory in galaxies]{The Contribution of Stars, Dust, Neutral Gas and Supermassive Black Holes in Galaxies to the Cosmic Baryon Inventory}

\author[J. C. J. D'Silva et al.]{
Jordan C. J. D'Silva,$^{1,2}$\thanks{E-mail: jordan.dsilva@research.uwa.edu.au} 
Simon P. Driver,$^{1}$
Aaron S. G. Robotham,$^{1}$
Andrew Battisti,$^{1,3}$ 
Elisabete da Cunha,$^{1}$ \newauthor
Luke J. M. Davies,$^{1}$ 
Stephen Eales,$^{4}$
Claudia del P. Lagos$^{1}$
\\
$^{1}$ International Centre for Radio Astronomy Research (ICRAR), University of Western Australia, Crawley, WA 6009, Australia\\
$^{2}$ Jodrell Bank Centre for Astrophysics, University of Manchester, Oxford Road, Manchester M13 9PL, UK\\
$^{3}$ Research School of Astronomy and Astrophysics, Australian National University, Cotter Road, Weston Creek, ACT 2611, Australia\\
$^{4}$ Cardiff Hub for Astrophysics Research and Technology, Cardiff University, The Parade, Cardiff CF24 3AA, UK
}

\date{Accepted XXX. Received YYY; in original form ZZZ}

\pubyear{\the\year{}}

\begin{document}
\label{firstpage}
\pagerange{\pageref{firstpage}--\pageref{lastpage}}
\maketitle

\begin{abstract}
We compute the cosmic stellar, dust and neutral gas mass history at $0<z\lesssim3$ using \textsc{ProSpect} spectral energy distribution modelling of $\approx 800 \, 000$ galaxies in the Galaxy and Mass Assembly (GAMA) survey and the Deep Extragalactic VIsible Legacy Survey (DEVILS). The cosmic dust mass history broadly follows the shape of the cosmic star formation history; though, the decline is slower, suggestive of a slowing rate of dust growth and destruction as the star formation declines past its peak at $z\approx 2$. Neutral gas masses were estimated by scaling the dust masses by the metallicity-dependent dust-to-gas ratio. The neutral gas mass density as traced by the dust is an average of \referee{$\approx 0.7$~dex} lower than that measured from $21$cm experiments, most likely due to differences in the spatial scales inhabited by dust and HI. Folding in measurements of the supermassive black hole mass density obtained previously with similar data and methods, we present a self-consistent census of the baryons confined to galaxies. Stars, neutral gas, SMBHs and dust contained within the optical radii of galaxies account for $\approx 5$ per cent of the baryons. Most of the remaining $\approx 95$ per cent of baryons must be ionised and dispersed throughout the interstellar, circumgalactic and intergalactic media within, around and between galaxies.
\end{abstract}

\begin{keywords}
galaxies: evolution -- galaxies: formation -- galaxies: general -- ISM: dust, extinction
\end{keywords}



\section{Introduction}\label{sec:intro}
Only $\approx 4$ per cent of the contents of the Universe are in the form of baryons \citep{aghanimPlanck2018Results2020c}. It is thought that the bulk of those baryons are ionised and in the intergalactic medium, with the neutral fraction only about 1 part in $10^{5}$ at $z\approx 5$ \citep{bosmanHydrogenReionizationEnds2022} following cosmic reionisation. The fraction of baryons inside of the galaxies is $\lesssim 10$ per cent \citep{graaffProbingMissingBaryons2019}. The stars and active galactic nuclei (AGNs) in galaxies are responsible for the second largest repository of extragalactic photons incident on the Earth after the cosmic microwave background radiation, radiating especially in the radio, optical and infrared wavelengths \citep{driverMeasurementsExtragalacticBackground2016a,koushanGAMADEVILSConstraining2021,tompkinsCosmicRadioBackground2023,chiangCosmicInfraredBackground2025a}. So, quantifying the baryon budget in galaxies is of great interest, particularly in connection to the astrophysical processes of star formation, chemical enrichment and the growth of supermassive black holes (SMBHs). 

Since $z\approx 2$, the cosmic star formation history has declined by a factor of $\approx 10$ on account of the depletion and consumption of their gas reservoirs \citep{madauCosmicStarFormationHistory2014,perouxCosmicBaryonMetal2020,walterEvolutionBaryonsAssociated2020}. The cosmic stellar mass history (CSMH) continues to steadily increase as the stellar mass cumulatively adds up with star formation \citep{wrightGAMAG10COSMOS3DHST2018,driverGAMAG10COSMOS3DHST2018a,thorneDeepExtragalacticVIsible2021,shuntovCOSMOSWebStellarMass2025}. A product of stellar mass assembly is chemical enrichment and the formation of dust grains. Asymptotic giant branch stars produce dust grains in their circumstellar envelopes that can be released into the interstellar medium (ISM) via stellar winds \citep[e.g.,][]{venturaDustAsymptoticGiant2014,hofnerMassLossStars2018,hofnerExplainingWindsAGB2020}. Dust also enters the ISM through supernovae explosions \citep[e.g.,][]{nomotoNucleosynthesisYieldsCorecollapse2006,sarangiDustSupernovaeSupernova2018,marassiSupernovaDustYields2019,bocchioDustGrainsHeart2016,schneiderFormationCosmicEvolution2024}, and can continue to condense in the remnants \citep{woodenAirborneSpectrophotometrySN1993,matsuuraHerschelDetectsMassive2011}. Dust growth may be tempered when, for example, the grains are destroyed in the tumultuous ISM or through astration where the dust is entombed in newly formed stars \citep{drainePhysicsDustGrains1979,draineDestructionMechanismsInterstellar1979,jonesGrainDestructionShocks1994,jonesGrainShatteringShocks1996,micelottaDustSupernovaeSupernova2018,dayalALMAREBELSSurvey2022}.

Dust is a critical component of galaxies especially due to its effect on their light distributions. Photons with wavelengths largely between the far ultraviolet (FUV) to optical from stars and AGNs are attenuated and the absorbed energy is reradiated in the infrared (IR), mid infrared (MIR) and far infrared (FIR) \citep[e.g.,][]{trumplerAbsorptionLightGalactic1930,draineInterstellarDustGrains2003}. The energy balance argument means that the reradiated emission is a tracer of star formation as the young stars heat up the dust \citep{kennicuttjrStarFormationMilky2012,daviesGAMAHATLASMetaanalysis2016}. The cosmic dust mass history (CDMH) encodes the balance between dust growth and destruction that, consequently, contains clues for recipes of chemical enrichment and feedback in galaxy formation models \citep{somervillePhysicalModelsGalaxy2015a,trayfordModellingEvolutionInfluence2026}. 

Dust grains are known to be catalysts for molecular hydrogen formation as $\mathrm{H_{2}}$ molecules condense on their surfaces. \citep{wakelamH2FormationInterstellar2017}. This is particularly useful because H$_{2}$ is notoriously hard to observe in the normal conditions of the giant molecular clouds where the temperatures are generally too low to excite the molecules. In lieu of viable direct H$_{2}$ measurements, tracers are observed instead, such as CO, CI and [CII] \citep{weissGasDustCloverleaf2003,papadopoulosEmissionUltraluminousInfrared2004, bolattoCOtoH2ConversionFactor2013, zanellaIiEmissionMolecular2018,maddenTracingTotalMolecular2020,dunneDustCOCrosscalibration2022}. Dust is also a tracer of H$_{2}$ to complement the tracers based on carbon \citep{dunneDustCOCrosscalibration2022}. 

In this paper, we conduct a census of the baryons confined to galaxies, taking into account the contribution from, in particular, the dust and the neutral gas traced by dust. We do this with spectral energy distribution (SED) fitting, building on from a similar census of stars and dust undertaken by \citet{driverGAMAG10COSMOS3DHST2018a} who used SED fitting on $\sim 10^{5}$ galaxies. 

The novelty of this work is to compute the redshift evolution of the dust and neutral gas content, and its connection to most of the remaining baryons in galaxies, in a self-consistent way. Specifically, we will use the homogeneous data and SED fitting results at $0<z<3$ from the Galaxy And Mass Assembly \citep[GAMA,][]{driverGalaxyMassAssembly2011c,driverGalaxyMassAssembly2022b} survey and the Deep Extragalactic VIsible Legacy Survey (DEVILS, \citealt{daviesDeepExtragalacticVIsible2018b}; \citealt{daviesDeepExtragalacticVIsible2025}). A critical advantage of this data set is that the evolution of cosmic dust and neutral gas is then easily and directly comparable to the volume-averaged evolution of stellar mass, star formation rate and SMBH growth also inferred with GAMA and DEVILS. Weighing the baryons in this manner means that the mass budget in galaxies can be easily combined with the equivalent mass budget in the cosmic web to account for most, if not all, of the visible matter in the entire Universe \citep[e.g.,][]{macquartCensusBaryonsUniverse2020,connorGasrichCosmicWeb2025}.

The outline of the paper is as follows. The data is described in Section~\ref{sec:data}. The main results are presented in Section~\ref{sec:fitting_smf}, \ref{sec:cdmh} and \ref{sec:baryonInventory}. Our final conclusions and summary of the work is in Section~\ref{sec:conclusion}. We use \textit{Planck} $\Lambda$CDM cosmology with $\mathrm{H_{0} = 68.4 \, km \, s^{-1} \, Mpc^{-1}}$, $\mathrm{\Omega_{\Lambda}} = 0.699$ and $\mathrm{\Omega_{M}} = 0.301$ \citep{aghanimPlanck2018Results2020c}.

\section{Data}\label{sec:data}
\subsection{Photometry}\label{sec:photometry}
GAMA and DEVILS are vast volume-complete, spectro-photometric surveys, covering $0<z<4$ with high completeness. Both surveys are complemented by FUV-FIR photometric measurements as processed by the source finding software, \textsc{ProFound} \citep{robothamProFoundSourceExtraction2018b}. The details of the photometric extraction can be found in \citet{bellstedtGalaxyMassAssembly2020a} for GAMA and \citet{daviesDeepExtragalacticVIsible2021} for DEVILS. For both GAMA and DEVILS the source detection was first performed on the $r+Z$ (GAMA) and $Y+J+H$ (DEVILS) bands to extract UV-optical-NIR photometry. MIR-FIR photometry for both surveys was extracted using a point-spread-function (PSF) aperture approach to account for the differences in both the spatial resolution and sensitivity of the different images. The source positions for the PSF photometry were obtained from the optically selected catalogue. 

For DEVILS, PSF photometry was not sought for every source in the $Y+J+H$ selected catalogue because the $Y-$flux is not well correlated with the expected emission at longer wavelengths and the lower resolution of the MIR-FIR data meant that some sources will be confused with multiple input galaxies occupying the same resolution element \citep{daviesDeepExtragalacticVIsible2021}. PSF fitting was first performed on the $24\mu\mathrm{m}$ Multiband Imaging Photometer for Spitzer (MIPS24) imaging \citep{sandersSCOSMOSSpitzerLegacy2007a} using the coordinates of a sample of $Y<21.2$~mag objects. The PSF fitting routine also extracted sources that were bright in the MIPS24 imaging but had no clear UV-MIR counterpart in the initial $Y<21.2$~mag sample. For these sources, a likely counterpart was assigned by cross matching the source coordinates against the positions in the full UV-MIR catalogue (including $Y>21.2$~mag objects). The $Y-24\mu\mathrm{m}$ colour was also cross matched against $Y-24\mu\mathrm{m \approx 3 \, mag}$, which was identified as the typical colour for the of the $Y<21.2$~mag sample. This process of matching will fail for sources with atypical $Y-24\mu\mathrm{m}$ colours. \referee{However, the number of sources detected in the MIPS24 imaging but with $Y>21.2$~mag (i.e., no clear optical counterpart that would require matching back to the full sample) is $8830$ or $\approx 2$ per cent of the total DEVILS sample used here. This means that even when the matching may have failed for atypical sources, the level of bias introduced is likely small.}

The source positions from the MIPS24 sample were then used for further PSF photometry in the MIPS70 band and the \textit{Herschel} $P100,P160,S250,S350,S500$ bands \citep{lutzPACSEvolutionaryProbe2011a,oliverHerschelMultitieredExtragalactic2012a}.
\begin{figure}
    \centering
    \includegraphics[width=\linewidth]{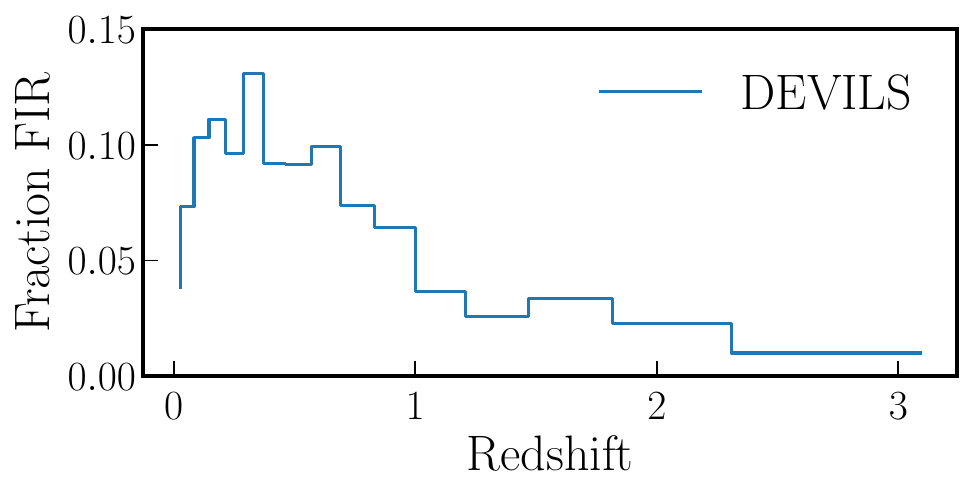}
    \caption{Fraction of sources with FIR photometric observations in the DEVILS sample.}
    \label{fig:devils_fir_frac}
\end{figure}
Figure~\ref{fig:devils_fir_frac} shows that the fraction of MIR-FIR observations in the DEVILS sample is $\approx 10$ per cent at $z\approx 0.3$ and decreases to $\approx 1$ percent at $z>2$. 

No attempt was made to measure the FIR photometry for objects that did not satisfy the above requirements (i.e., $Y>21.2$~mag sources without any MIPS24 counterpart), and the fluxes have no value. A small fraction of FIR objects were detected with no clear counterpart in either the optical or MIPS24 bands. These objects were not included in the DEVILS SED fitting results because they were not associated with a DEVILS target.

A similar PSF extraction was performed on the GAMA imaging but, unlike DEVILS, the imaging was wider and shallower meaning that FIR photometry could be extracted for essentially all optically detected sources \citep{thorneDeepExtragalacticVIsible2021}. We only use the D10 sub-survey of DEVILS in the COSMOS field because of its spectroscopic and photometric redshift completeness that is complemented by previous surveys \citep[e.g., zCOSMOS,][]{lillyZCOSMOS10kBRIGHTSPECTROSCOPIC2009} and to aid in comparison with previous works \citep{thorneDeepExtragalacticVIsible2021,dsilvaGAMADEVILSCosmic2023}.

\subsection{SED fits}\label{sec:sed-fits}
The SEDs were fit using \textsc{ProSpect} \citep{robothamProSpectGeneratingSpectral2020}, in work first presented in \citet{bellstedtGalaxyMassAssembly2020b} for GAMA and \citet{thorneDeepExtragalacticVIsible2021,thorneDeepExtragalacticVIsible2022} in the D10-COSMOS field of DEVILS. These previous SED fits of $\approx 800 \, 000$ galaxies between GAMA and DEVILS-D10 are used in this work. \textsc{ProSpect} models the FUV-FIR emission of galaxies under the assumption of energy balance. 

The intrinsic stellar emission is modelled with the \citet{bruzualStellarPopulationSynthesis2003} stellar population synthesis library, the \citet{chabrierGalacticStellarSubstellar2003} initial mass function (IMF) and a flexible skewed-normal, parametric star formation history. 

The intrinsic star light is then attenuated using the \citet{charlotSimpleModelAbsorption2000} model. Attenuation as a function of wavelength is calculated as \begin{equation}
    F_{\lambda}^{\mathrm{observed}} = F_{\lambda}^{\mathrm{intrinsic}} \, e^{-\tau_{V}(\lambda / \lambda_{0})^{-0.7}},
\end{equation}
where \referee{$F_{\lambda}^{\mathrm{observed}}$ is the observed flux, $F_{\lambda}^{\mathrm{intrinsic}}$ is the intrinsic flux}, $\lambda_{0} = \SI{5500}{\angstrom}$ and $\tau_{V}$ is the $V$-band normalising coefficient of the optical depth. This is a two component model where light is attenuated both by stellar birth clouds and the ISM, however only stars younger than $10 \, \mathrm{Myr}$ will have their light affected by the birth clouds. Hence, in \textsc{ProSpect} the two parameters that control the attenuation are $\tau_{V}^{\mathrm{BC}}$ and $\tau_{V}^{\mathrm{ISM}}$. 

The metallicity is implemented as an evolutionary model that scales linearly with the stellar mass growth and chemical enrichment. The initial metallicity is assumed to be $Z=10^{-4}$ that is the lowest metallicity template in the \citet{bruzualStellarPopulationSynthesis2003} stellar population synthesis library. The parameter to be fitted is the final metallicity at the epoch of observation ($\mathrm{Z_{final}}$) that represents the metallicity of the gas from which the final generation of stars formed. $\mathrm{Z_{final}}$ for a fitted galaxy is closest to the gas-phase metallicity measured from spectral lines, unless the gas content of the galaxy changed significantly after the last epoch of star formation. For example, if the galaxy continued to accrete metal-poor gas then $\mathrm{Z_{final}}$ would be higher than the gas-phase metallicity \citep{thorneDEVILSCosmicEvolution2022}. \referee{$\mathrm{Z_{final}}$ can be interpreted as the metallicity in HII regions. Observations indicate the metallicity measured from both photoionised gas in HII regions and neutral gas in the wider ISM are similar \citep[e.g.,][]{pettiniNewObservationsInterstellar2002,friisWarmExcitedMolecular2015a,hernandezFirstCospatialComparison2021}.}

The attenuated light is then assumed to be reradiated in the IR-FIR as per the \citet{daleTwoparameterModelInfrared2014} dust emission models. These models were computed by combining the emission curves of large dust grains with particle sizes $\sim 100 \, \mathrm{nm}$, small grains with sizes $\sim 10 \, \mathrm{nm}$ and poly-aromatic hydrocarbon (PAH) molecules with sizes $\sim 1 \, \mathrm{nm}$ \citep{desertInterstellarDustModels1990} after exposing them to a variety of radiation fields with various levels of intensity, $U$ \citep{daleInfraredSpectralEnergy2001,daleInfraredSpectralEnergy2002}. The \citet{daleTwoparameterModelInfrared2014} models envisage the total dust emission of a galaxy as an ensemble of local dust emission SEDs exposed to radiation fields with a range of intensities, $ 0.3 \leq U \leq 10^{5}$, where $U=1$ is the intensity in the solar neighbourhood. 

In the \citet{daleTwoparameterModelInfrared2014} models, the distribution of the strengths of the local radiation fields is controlled by the parameter $\alpha$, such that 
\begin{equation}
    \label{eq:dale-alpha}
    \mathrm{dM_{dust}} \propto U^{-\alpha}\mathrm{d}U,
\end{equation}
where $\mathrm{dM_{dust}}$ are the local dust masses exposed to the radiation fields. Small values of $\alpha$ produce lower dust masses than larger values because hot, large dust grains dominate the emission \citep{daleInfraredSpectralEnergy2002}. \textsc{ProSpect} considers a superposition of two \citet{daleTwoparameterModelInfrared2014} models with $\mathrm{\alpha_{BC}}$ and $\mathrm{\alpha_{ISM}}$ to account for dust in the stellar birth clouds and the ISM. 

The presence of an AGN can affect the resulting astrophysical quantities. \textsc{ProSpect} has the ability to fit an AGN component using the \citet{fritz06agnmodel} model, the UV-optical emission of which is also attenuated by the dust screen of the ISM. We denote the sets of SED fits with an AGN component included as $\mathrm{Pro-Stellar+AGN}$ while those with only stellar emission we call $\mathrm{Pro-Stellar}$. $\mathrm{Pro-Stellar}$ has fewer parameters \citep[see tab.2 in][]{thorneDeepExtragalacticVIsible2021} than $\mathrm{Pro-Stellar+AGN}$ \cite[see Tab. 1 in][]{thorneDeepExtragalacticVIsible2022} because there is no significant AGN component to be modelled. The optimal solution likely resides somewhere between these two that we refer to as $\mathrm{Pro-Hybrid}$. Being guided by the principle of parsimony, $\mathrm{Pro-Hybrid}$ was obtained by first preferring $\mathrm{Pro-Stellar}$ for every SED fit and only swapping to $\mathrm{Pro-Stellar+AGN}$ when the likelihood of the $\mathrm{Pro-Stellar+AGN}$ was preferred and the AGN contribution to the rest-frame fitted SED at $5-20\, \mu \mathrm{m}$ was more than $10$ per cent. This metric and threshold for significant AGN is the same as that outlined by \citet{daleTwoparameterModelInfrared2014,thorneDeepExtragalacticVIsible2022}. 

\section{Calculating the dust mass}
As we wish to derive the contribution of dust to the cosmic baryon inventory, it is important to discuss how the dust mass is inferred from SED fitting with \textsc{ProSpect}.

In principle, the dust mass can be found by integrating Equation~\ref{eq:dale-alpha} over $U$ once $\alpha$ has been fitted and the proportionality constant, which connects the strength of the radiation field to the dust emission, is known. Alternatively and more simply, the \citet{daleTwoparameterModelInfrared2014} models calculate the IR-FIR emission for $1 \, \mathrm{M_{\odot}}$ of dust and so the total dust mass of a galaxy is found by comparing the absorbed luminosity as per the \citet{charlotSimpleModelAbsorption2000} model with the bolometric luminosity of the fitted \citet{daleTwoparameterModelInfrared2014} model (more discussion on this is in Section~\ref{sec:modellingAssumptions}). In terms of the dust mass, the salient parameters are $\tau_{V}^{\mathrm{BC}}$, $\tau_{V}^{\mathrm{ISM}}$, $\mathrm{\alpha_{BC}}$ and $\mathrm{\alpha_{ISM}}$ that together control the dust attenuation \citep{charlotSimpleModelAbsorption2000} and re-emission \citep{daleTwoparameterModelInfrared2014} for the two component dust model. 

\subsection{Dust masses without FIR photometry}\label{sec:dustMassNoFIR}
The same dust that is attenuating the blue star light is re-radiating at redder wavelengths in the FIR. Hence, the dust mass is inferred from assumptions about the absorbed luminosity and the shape of the thermal emission spectrum and the dust temperature.
\begin{figure}
    \centering
    \includegraphics[width=\linewidth]{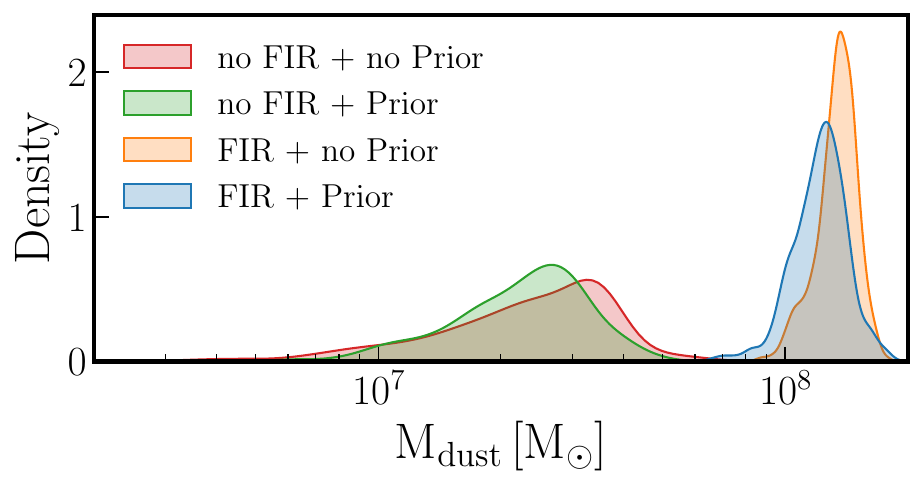}
    \caption{Posterior distributions of the dust mass from \textsc{ProSpect} fits to the inverse-variance weighted stack of a sample of $z\approx 0.05$ galaxies with $S/N > 5$ FIR \textit{Herschel} $P100,P160,S250,S350,S500$ photometric measurements. The blue curve shows the results when including the FIR measurements and a Gaussian prior on $\alpha$ with mean $\mu=2$ and dispersion $\sigma=1$. The orange curve shows the results including when including the FIR measurements but no prior. The green curve shows the results when excluding the FIR measurements and using the prior. The red curve shows the results when excluding both the FIR measurements and the prior.}
    \label{fig:fir_dust_mass_errs}
\end{figure}
DEVILS becomes more and more important at higher and higher redshifts. 
The fraction of sources with FIR measurements diminishes with redshift (as shown in. Figure~\ref{fig:devils_fir_frac}), which has implications for the constraints of dust temperature and, therefore, dust mass. To explore this, the inverse-variance weighted stacked SED of a sample of $z\approx 0.05$ GAMA galaxies with $S/N > 5$ photometry in all of the FIR \textit{Herschel} $P100,P160,S250,S350,S500$ photometric bands was obtained and the SED was fit with and without FIR photometric measurements. The posterior distributions of the dust masses are shown in Figure~\ref{fig:fir_dust_mass_errs}.

It is clear that the absence of the FIR introduces significant uncertainty on the dust mass, as evident from the difference in the variance of the posterior distributions, because of the sensitivity on the dust temperature. The \textsc{ProSpect} fits therefore used a Gaussian prior on each of the $\alpha$ parameters with dispersion $\sigma=1$ and mean $\mu=2$, which is the middle value of the range of $\alpha \in [0,4]$ considered for SED fitting \citep{thorneDeepExtragalacticVIsible2021,thorneDeepExtragalacticVIsible2022}. The inclusion of the FIR also systematically produces $\approx 0.7$~dex more dust mass because the fit prefers $\alpha \approx 3$ and therefore cooler dust than for $\alpha = 2$. This is also why including this prior on $\alpha$ results in a less sharply peaked posterior distribution. 

There is evidence that the dust temperature increases with redshift \citep[e.g.,][]{magnelliEvolutionDustTemperatures2014, liangDustTemperaturesHighredshift2019, chiangCosmicInfraredBackground2025a}, meaning that $\alpha \approx 3$ is unsurprising for this low redshift sample. The temperature from Wien's law and the peak of the fitted FIR emission was $\approx 25$~K for $\alpha=3$ and $\approx 68$~K for $\alpha = 2$. This test shows, as a worst case scenario, that without FIR measurements the dust mass is highly sensitive to the assumption about the dust temperature. The Gaussian prior on $\alpha$ was set under the assumption that the dust is hotter at high redshift than low, especially in regard for the DEVILS galaxies. Fortunately, the situation at high redshift can be improved with joint observations from, for example, the \textit{James Webb Space Telescope} and the \textit{Atacama Large Millimeter Array} \citep[e.g.,][]{caseyDustLittleRed2024,cieslaDustEmissionBulk2025,heintzInefficientDustProduction2025,algeraREBELSIFUDustBuildup2026}. 

\subsection{Dust mass modelling assumptions}\label{sec:modellingAssumptions}
\citet{driverGAMAG10COSMOS3DHST2018a,dunneHerschelATLASRapidEvolution2011,pozziDustMassFunction2020,beestonConfirmingEvolutionDust2024,bertaPanchromaticViewN2CLS2025} all used the SED fitting code MAGPHYS \citep{dacunhaSimpleModelInterpret2008} to infer dust masses for the purpose of calculating the CDMH. MAGPHYS characterises the total IR-FIR emission as an ensemble of grey body radiators where the relative contributions and their temperatures are fitted parameters. 

Both MAGPHYS and \textsc{ProSpect} require robust measurements of the FIR emission, lest the dust masses be highly uncertain, as discussed in Section~\ref{sec:dustMassNoFIR}. To explore the intrinsic differences between the two SED fitting codes, we isolated a sample of $218$ galaxies with $S/N > 5$ in the \textit{Herschel} $P100,P160,S250,S350,S500$ photometric bands in both the \textsc{ProFound} catalogue \citep{bellstedtGalaxyMassAssembly2020a} used for the \textsc{ProSpect} SED fitting of this work and the LAMBDAR catalogue \citep{wrightGalaxyMassAssembly2016a} used for the MAGPHYS fitting in \citet{driverGAMAG10COSMOS3DHST2018a}.

\begin{figure}
    \centering
    \includegraphics[width=\columnwidth]{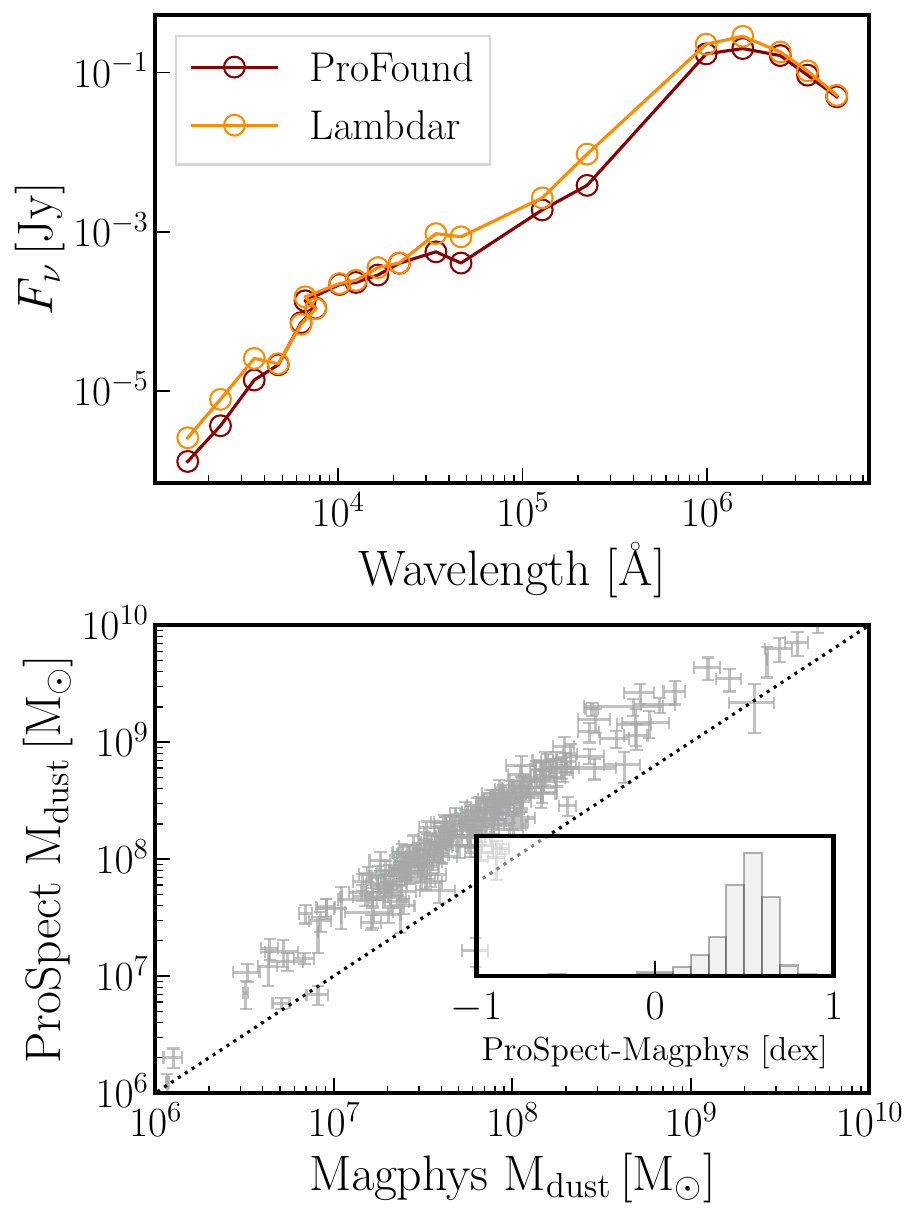}
    \caption{\textit{Top:} inverse-variance weighted stacked SEDs of the $218$ galaxies with $S/N > 5$ in the \textit{Herschel} $P100,P160,S250,S350,S500$ photometric bands from both \textsc{ProFound} and LAMBDAR photometry. \textit{Bottom:} \textsc{ProSpect} dust masses and $1\sigma$ uncertainties against MAGPHYS. The grey dotted line is the equality line. The inset panel is the histogram of the dust mass differences between \textsc{ProSpect} and MAGPHYS}
    \label{fig:magphys-prospect}
\end{figure}
Figure~\ref{fig:magphys-prospect} shows the comparison between \textsc{ProSpect} and MAGPHYS for this robust sample. The top panel shows an inverse-variance weighted stack of the SEDs of the robust sample, showing relative consistency between the two photometry tools. The bottom panel shows the histogram of the differences between the dust masses. \textsc{ProSpect} infers $\approx 0.5$~dex higher dust masses than MAGPHYS. \referee{As this subsample is selected for} robust constraints on the FIR emission, the disparity is mostly due to differences in the modelling assumptions between codes and not from differences in the constraints of the FIR emission. 

The MAGPHYS model includes contributions from birth clouds, inhabited by stars less than $10$~Myr old, and the ISM. Each of those components contains a further contribution from $\sim 100$~K dust in the MIR, PAH emission using the template of \citet{maddenISMPropertiesLowmetallicity2006}, warm dust of $\approx 50$~K and cold dust of $\approx 20$~K. These components are assumed to have fractional contributions to the total IR-FIR emission such that the total of each contribution must sum to $1$. For further details on the dust model in MAGPHYS, the reader is directed to \citet{dacunhaSimpleModelInterpret2008} and in particular Eqs. 13-22. The dust masses on the other hand are calculated only from the contributions of the warm component in the birth clouds and the warm and cold components in the ISM, assuming that they are grey body radiators \citep{hildebrandDeterminationCloudMasses1983a}. The contribution from the MIR and PAH emission is not explicitly accounted for because, as argued in \citet{dacunhaSimpleModelInterpret2008}, PAHs only contribute a few percent of the total dust mass \citep[e.g.,][]{draineDustMassesPAH2007,remy-ruyerLinkingDustEmission2015,anianoModelingDustStarlight2020,liSpitzersPerspectivePolycyclic2020,chastenetPHANGSJWSTFirst2023,shivaeiNewCensusDust2024,sutterFractionDustMass2024}, and the final dust mass in MAGPHYS is therefore simply scaled up by 10 per cent.

The \citet{daleTwoparameterModelInfrared2014} models provide the dust emission templates in units of $\mathrm{[W/H_{atom}]}$. Therefore, in \textsc{ProSpect} the luminosity for a $1 \, \mathrm{M_{\odot}}$ of dust is calculated as
\begin{equation}
    \label{eq:ML}
    (M/L_{i})^{-1} = \int^{\infty}_{0} \frac{f_{i}(\lambda)/\lambda}{ \mathrm{L_{\odot} \, m_{H}/M_{\odot} \, DTH} } \, d\lambda,
\end{equation}
and the total dust mass of a galaxy fitted by a given \citet{daleTwoparameterModelInfrared2014} template, $f_{i}(\lambda)$, is found by dividing its total absorbed luminosity from the \citet{charlotSimpleModelAbsorption2000} model by $(M/L_{i})^{-1}$. A key assumption in \textsc{ProSpect} is a constant dust-to-hydrogen mass ratio (DTH) of $\mathrm{DTH = 0.0073}$ that is the expected value for solar metallicity in the Milky Way \citep{draineDustMassesPAH2007}. The assumption is that the DTH is the same for all dust species that emit across the breadth of the electromagnetic spectrum meaning that the mass-to-light ratio for all dust species are the same. 

However, PAH molecules that emit in predominantly the MIR only contribute a few per cent of the total mass in dust despite their significant contribution to the SED \citep{draineDustMassesPAH2007,remy-ruyerLinkingDustEmission2015,anianoModelingDustStarlight2020,liSpitzersPerspectivePolycyclic2020,chastenetPHANGSJWSTFirst2023,shivaeiNewCensusDust2024,sutterFractionDustMass2024}. The \citet{daleTwoparameterModelInfrared2014} models were originally constructed from grain mixtures of very small grains, PAHs and big grains, before being combined into a total emission model. The big grains dominate the mass and the FIR emission, whereas the remaining dust species contribute less mass and mostly emit in the MIR \citep{daleInfraredSpectralEnergy2002,desertInterstellarDustModels1990}. Because of this we recalculated the \textsc{ProSpect} dust masses by weighting the DTH as a function of wavelength to account for the difference in the mass contribution between grain species. 

\begin{figure}
    \centering
    \includegraphics[width=\columnwidth]{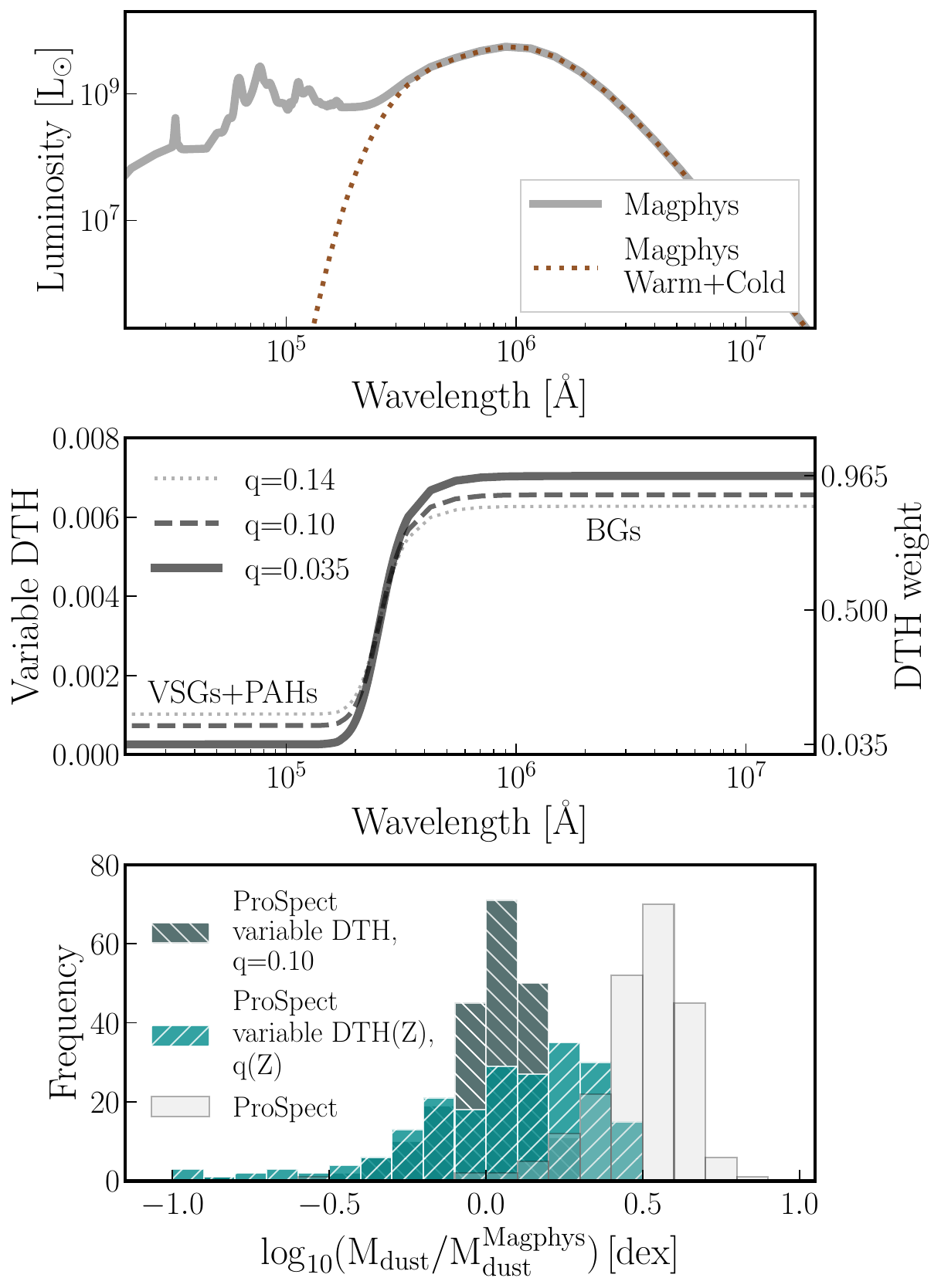}
    \caption{\textit{Top:} MAGPHYS standard SED model. The thick grey line is for the total luminosity while the brown, dotted line is only the mass contributing part of the SED. \textit{Middle:} variable DTH as a function of wavelength. The left axis denotes the DTH per wavelength while the right axis denotes the factor that $\mathrm{DTH=0.0073}$ is multiplied by. The regimes at which very small grains (VSGs), PAHs and big grains (BGs) dominate the emission are noted. \referee{We show three different lines, as indicated in the legend, to indicate different fractional contributions to the dust mass, $q$, from very small grains and PAHs.} \textit{Bottom:} histogram comparison of dust masses between MAGPHYS. The light grey histogram is the same comparison in Figure~\ref{fig:magphys-prospect}. The dark histogram is the comparison against MAGPHYS using the updated dust masses and the variable DTH with $q=0.1$. \referee{The cyan histogram shows the comparison when using the metallicity-dependent DTH from \citet{remy-ruyerGastodustMassRatios2014a} and the metallicity-dependent $q(Z)$ from \citet{shivaeiNewCensusDust2024}.}}
    \label{fig:variable_dth}
\end{figure}
Figure~\ref{fig:variable_dth} shows the calculation of our variable DTH, where we took the ratio of the standard MAGPHYS template of the mass contributing SED to the total SED as the weighting to apply to $\mathrm{DTH=0.0073}$. It can be seen that this variable DTH is close in form to a piecewise function,
\begin{equation}
\mathrm{DTH}(\lambda) \propto
\begin{cases} 
    q & \lambda \lesssim 2\times10^{5}\,\SI{}{\angstrom} \\
    1-q & \lambda \gtrsim 2\times10^{5}\,\SI{}{\angstrom},
\end{cases}
\end{equation}
where $q$ is the fraction of mass in PAHs$+$very small grains and the proportionality is the total DTH. Using $q=0.10$ the refined dust masses with the updated variable DTH agree with the dust masses from MAGPHYS. The reader is reminded that MAGPHYS includes a $10$ per cent scaling. On average, the dust masses are lowered by $\approx 3.1$ times; hence, this is expected to be \referee{an important factor driving the differences between \textsc{ProSpect} and MAGPHYS dust masses}.

\referee{It must be stressed however that this scaling is sensitive to the value of $q$. For example, $q=0.14$ results in $\approx 2.5 $ times lower dust masses. This value of $q=0.14$ is the same that was used for the grain mixture in the construction of the \citet{desertInterstellarDustModels1990, daleInfraredSpectralEnergy2002} models, on which the \citet{daleTwoparameterModelInfrared2014} were based. A value of $q=0.035$ results in $\approx 6.8$ times lower dust masses. These numbers are based on the solar DTH$=0.0073$.}

\subsection{Metallicity-dependent DTH and $q$}
In addition to a variable DTH per dust species, there is convincing evidence that the DTH changes with the gas-phase metallicity, with metal poorer systems having lower DTH \citep[e.g.,][]{casasolaISMScalingRelations2020,perouxCosmicBaryonMetal2020,gallianoNearbyGalaxyPerspective2021,parkSpatiallyResolvedRelation2024}. The dust masses from GAMA and DEVILS were corrected for this effect by rescaling the DTH with the broken power law fit to the DTH-$Z$ from \citet[][$X_{\mathrm{CO,Z}}$, see their Tab. 1]{remy-ruyerGastodustMassRatios2014a} as a function of the $Z_{\mathrm{final}}$ fitted value from \textsc{ProSpect}. \referee{The \citet{remy-ruyerGastodustMassRatios2014a} relation includes both atomic and molecular hydrogen in the denominator of the DTH. At solar metallicity \citep[$Z_{\odot} = 0.014$,][]{asplundChemicalMakeupSun2021a} the \citet{remy-ruyerGastodustMassRatios2014a} DTH is $0.0086$.}

\referee{180 out of the 218 galaxies in the sample used in Figure~\ref{fig:variable_dth} have super-solar, $Z_{\mathrm{final}}>Z_{\odot}$, metallicities and the resulting DTH from the \citet{remy-ruyerGastodustMassRatios2014a} relation can be a factor of $\approx 3$ higher than $\mathrm{DTH = 0.0073}$ and worsen the agreement between the dust masses obtained from MAGPHYS and \textsc{ProSpect}. This suggests that values of $q<0.10$ should be used in conjunction with the metallicity-dependent DTH. Observations indicate that very small grains and PAHs constitute $1-6$ per cent of the total dust mass and that the mass fraction changes with metallicity \citep{draineDustMassesPAH2007,remy-ruyerLinkingDustEmission2015,anianoModelingDustStarlight2020,liSpitzersPerspectivePolycyclic2020,chastenetPHANGSJWSTFirst2023,shivaeiNewCensusDust2024,sutterFractionDustMass2024}.}

\referee{We recalculatec the \textsc{ProSpect} dust masses of the 218 galaxies using the \citet{remy-ruyerGastodustMassRatios2014a} DTH-$Z$ relation and the $q(Z)$ relation from \citet{shivaeiNewCensusDust2024}. This $q(Z)$ is $1$~per cent at $Z \lesssim 0.3 \, Z_{\odot}$, linearly increases up to $Z \approx 0.5 \, Z_{\odot}$ and asymptotes to $3.5$~per cent at $Z \gtrsim 0.5 \, Z_{\odot}$ (most of the 218 galaxies will have $q=0.035$). The \textsc{ProSpect} dust masses calculated this way are on average $\approx 1.3$ times higher than MAGPHYS. The spread in the distribution reflects the spread in the metallicities.}

\referee{A caveat is that the DTH-$Z$ relation is itself uncertain. While \citet{remy-ruyerGastodustMassRatios2014a} report a broken-power law fit, other works report single power law fits \citep{deciaDustdepletionSequencesDamped2016,wisemanEvolutionDusttometalsRatio2017,devisSystematicMetallicityStudy2019,heintzCosmicBuildupDust2023,yatesImpactBinaryStars2024}. For reference, the DTH from the single power law fit by \citet[][PG16S, see their Tab. 4]{devisSystematicMetallicityStudy2019} results in a DTH ratio that is $\approx 10$ times higher than the \citet{remy-ruyerGastodustMassRatios2014a} relation to $Z=0.05$, $\approx 1.6$ times higher at solar metallicity and $\approx 2.2$ times higher at $Z=10^{-4}$. }

\referee{Exploring the differences between DTH$-Z$ in the literature is beyond the scope of this work; hence, for the remainder of the analysis we use dust masses computed with the metallicity-dependent DTH from \citet{remy-ruyerGastodustMassRatios2014a} combined with the $q(Z)$ from \citet{shivaeiNewCensusDust2024}, as discussed.}

\subsection{Additional dust mass caveats} \label{sec:caveats}
MAGPHYS and the models of \citet{daleTwoparameterModelInfrared2014} also differ in their implementation of the emissivity index, $\beta$, for the greybody FIR emission, described as,
\begin{equation}
    \mathrm{greybody_{\lambda}(\beta, T_{dust}) \propto \lambda^{-\beta}B_{\lambda}(T_{dust})},
    \label{eq:greybody}
\end{equation}
where $\mathrm{B_{\lambda}(T_{dust})}$ is the Planck function. MAGPHYS uses $\beta = 1.5$ for warm dust and $\beta = 2.0$ for cold dust. \citet{daleTwoparameterModelInfrared2014} connect the emissivity index to the strength of the radiation field, $U$, that is heating the ensemble of dust masses (see Equation~\ref{eq:dale-alpha}), 
\begin{equation}
    \beta = 2.5 + 0.4\log_{10}(U).
\end{equation}
The effect of this is similar to the implementation in MAGPHYS, with lower $U$ values, and hence colder dust, corresponding to larger $\beta$ values; and the range of assumed $U$ values corresponds to $0.5 \lesssim \beta \lesssim 3$.

To better understand what effect this has on the dust mass between the two models, we fitted single-temperature greybody functions to the \textit{Herschel} photometry of the sample of $218$ FIR detected galaxies that were used to compare MAGPHYS and \textsc{ProSpect} in Section~\ref{sec:modellingAssumptions}. 
\begin{figure}
    \centering
    \includegraphics[width=\linewidth]{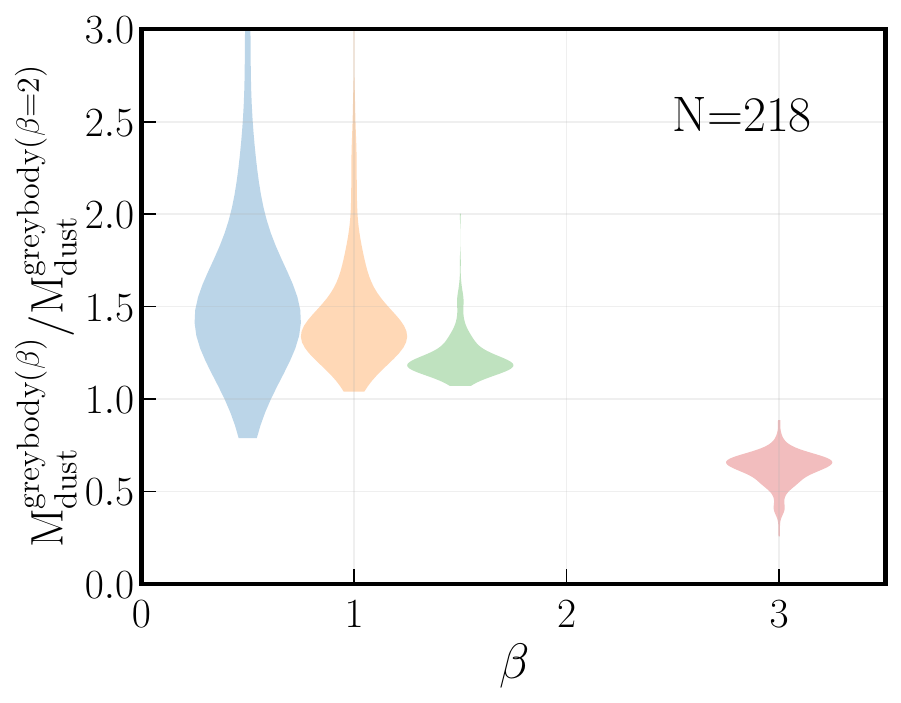}
    \caption{Comparison of the dust mass from fitting 218, FIR detected galaxies with single-temperature greybody functions. We show the distributions of the ratio of dust masses when varying the emissivity index, $\beta$, compared to a fixed value of $\beta=2$, which is similar to the assumed implementation in MAGPHYS.}
    \label{fig:greybodies}
\end{figure}
Figure~\ref{fig:greybodies} shows how different assumptions of $\beta$ compared to a fixed value of $\beta = 2$ affect the dust mass. In no case of $0.5 \lesssim \beta \lesssim 3$ (the extrema of the \citet{daleTwoparameterModelInfrared2014} models) is the average difference in the dust mass close to a factor of $3$. We, therefore, expect that the difference in $\beta$ between \citet{daleTwoparameterModelInfrared2014} and MAGPHYS cannot be the main driver of the dust mass difference. 

While we anticipate that the implementation of $\beta$ is not significantly different enough between the models to affect the dust mass, the opacity coefficient normalised at $850\mu\mathrm{m}$, $\kappa_{850\mu\mathrm{m}}$, could be important. $\kappa_{850\mu\mathrm{m}}$ is an essential parameter to obtain the absolute dust mass in MAPGHYS, and the assumption is that $\kappa_{850\mu\mathrm{m}} = 0.077 \, \mathrm{kg \, m^{-2}}$ for all galaxies \citep{dunneSCUBALocalUniverse2000}. The dust mass from the single-temperature greybody scales with $\kappa_{850\mu\mathrm{m}}$ such that the dust masses from MAGPHYS could have been $\approx 3$ times higher had $\kappa_{850\mu\mathrm{m}} =  \frac{1}{3} \, 0.077 \, \mathrm{kg \, m^{-2}}$. 

\citet{jamesSCUBAObservationsGalaxies2002} found that on average $\kappa_{850\mu\mathrm{m}} = 0.07 \pm 0.02 \, \mathrm{kg \, m^{-2}}$. More recently, \citet{clarkFirstMapsKd2019} found that $\kappa_{500\mu\mathrm{m}} = 0.11-0.25 \, \mathrm{kg \, m^{-2}}$ in M74 and $\kappa_{500\mu\mathrm{m}} = 0.15-0.80 \, \mathrm{kg \, m^{-2}}$ in M83 using spatially resolved maps, demonstrating that $\kappa_{500\mu\mathrm{m}}$ varies between these two galaxies and as a function of the local density of the ISM. 

The opacity at $500\mu\mathrm{m}$ can be connected to the opacity at $850\mu\mathrm{m}$ as $\kappa_{850\mu\mathrm{m}} = \kappa_{500\mu\mathrm{m}} (500/850)^{\beta}$, where $\beta$ is the emissivity index. From the ranges of $\kappa_{500\mu\mathrm{m}}$ from those two galaxies and values of $\beta = 1.5-2$, it is clear that on average $\frac{1}{3} \, 0.077 < \kappa_{850\mu\mathrm{m}} < \, 0.077 \, \mathrm{kg \, m^{-2}}$ for M74 and M83. So, despite \citet{clarkFirstMapsKd2019} finding lower $\kappa_{850\mu\mathrm{m}}$ values than $\kappa_{850\mu\mathrm{m}} = 0.077 \, \mathrm{kg \, m^{-2}}$, it is still not enough to entirely explain the factor of $\approx 3$ difference of dust mass between MAGPHYS and \textsc{ProSpect}. This is true provided that M74 and M83 are representative of all the galaxies used in this work. Ultimately, we cannot rule out an uncertain $\kappa_{850\mu\mathrm{m}}$ as causing differences in the dust masses between codes, but nevertheless maintain that a variable DTH \referee{(combined with the metallicity dependence)} is at least an improvement on what was originally assumed in \textsc{ProSpect}.

\section{Fitting the SMF}\label{sec:fitting_smf}
While the contribution of dust to the baryon inventory is the focus of this work, we first calculated the stellar mass contribution. The stellar mass distribution function (SMF) is calculated as,
\begin{equation}
    \phi(\mathrm{M_{\star}}) = \mathrm{\frac{dN}{dM_{\star}dV}},
\end{equation}
which is the histogram ($\mathrm{dN})$ of stellar masses ($\mathrm{dM_{\star}}$) per unit co-moving volume ($\mathrm{dV}$). The volume depends on the survey area and the limits of the redshift bins. We split our sample of GAMA and DEVILS into bins spanning $0.75$~Gyr of lookback time between 0 and 12 Gyr ago (corresponding to $z\approx 3.6$ in our cosmology), which was the same binning used by \citet{thorneDeepExtragalacticVIsible2021} for their calculations of the SMFs. 

We made this calculation separately for GAMA and DEVILS as they cover different areas on the sky, with GAMA covering $217.54 \, \mathrm{deg}^{2}$ and DEVILS-D10 covering $1.5 \, \mathrm{deg}^{2}$. Both of these surveys are volume complete up to a certain magnitude, meaning that the stellar mass bin at which the distribution turns down may be adopted as the incompleteness limit as per the Malmquist Bias \citep[e.g.,][]{weigelStellarMassFunctions2016}. We identified the maximum of the binned SMF for each GAMA and DEVILS and set that as the mass limit. The SMFs for both GAMA and DEVILS were then combined for a single SMF, leveraging the superior depth of DEVILS for the less massive end and the superior spatial coverage of GAMA for the massive end. This was essentially the same strategy used by \citet{driverGAMAG10COSMOS3DHST2018a} to compute the SMF by combining different surveys. \citet{thorneDeepExtragalacticVIsible2021} had previously presented SMFs at $0<z<4$ using \textsc{ProSpect} fits of DEVILS galaxies, and so the main improvement upon that work presented here is the inclusion of GAMA. 

When the slope of the SMF is steep, the uncertainties on the stellar masses can systematically change the shape of the intrinsic SMF compared to what is observed \citep[i.e., Eddington bias,][]{obreschkowEddingtonsDemonInferring2018}. The uncertainties on the SED fitted stellar masses were propagated onto the SMF through a Monte-Carlo experiment by varying their values within their $1\sigma$ uncertainties and recomputing the binned SMF 1000 times, and this was combined in quadrature with the Poisson uncertainties of the SMF. A final error floor of $10$ per cent on the SMF binned quantities was assumed to, at the very least, accommodate remaining systematic uncertainties. 

\begin{figure*}
    \centering
    \includegraphics[width=\linewidth]{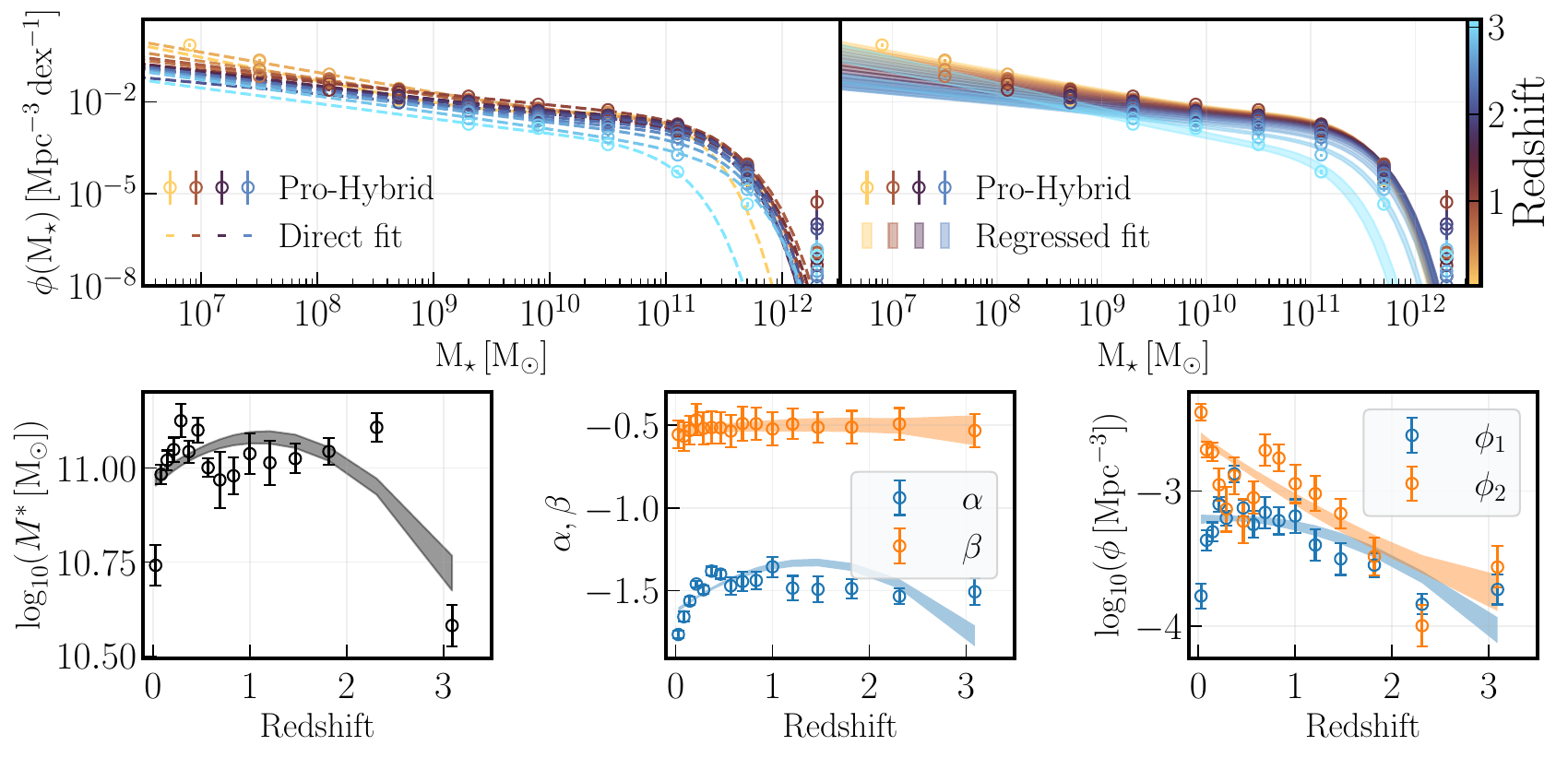}
    \caption{\textit{Top left:} SMFs (points with $1\sigma$ uncertainties, in some bins the size of the error bar is smaller than the points) and double Schechter function fits (dashed lines) from $z\approx 0$ (yellow-red) to $z\approx 3$ (blue). \textit{Bottom:} redshift evolution and $1\sigma$ uncertainties of $\log_{10}(M^{*})$ (left), $\alpha,\beta$ (middle) and $\phi_{1},\phi_{2}$ (right) as indicated in the legend. The shaded lines in all plots are the fits and $1\sigma$ uncertainties to the evolution which are quadratic in redshift. \textit{Top right:} the regressed SMFs from the quadratic fits to the parameter evolution. The width of the lines is the $1\sigma$ spread on the regressed SMFs. The colour scheme is the same as the top left panel. The fit parameters are presented in Table~\ref{tab:smf_par} and Table~\ref{tab:smf_par_evol}.}
    \label{fig:smf_regressed}
\end{figure*}
With the sets of binned stellar mass distributions and corresponding uncertainties across the 16 bins of redshift at $0<z\lesssim3$, we fitted the SMF with double Schechter functions, which was the same function used in \citet{thorneDeepExtragalacticVIsible2021}. The function is:
\begin{equation}
    \mathrm{\phi(M_{\star})} = \mathrm{\ln(10)} \, e^{-\mu} \, \left (  \phi_{1} \, \mu^{\alpha + 1} + \phi_{2} \, \mu^{\beta + 1} \right ),
    \label{eq:doubleSchechter}
\end{equation}
where $\mu = 10^{\mathcal{M}} / 10^{M^{*}}$ and $\mathcal{M} = \mathrm{\log_{10}(M_{\star}/M_{\odot})}$. The fitting software \textsc{Highlander}\footnote{\url{https://github.com/asgr/Highlander}}, which combines genetic optimisation and Markov-Chain-Monte-Carlo, was employed to find the maximum of the likelihood and the optimal solution of the parameters of the double Schechter functions. The top left panel of Figure~\ref{fig:smf_regressed} shows our fitted SMFs in each of the 16 redshift bins. 

Because of large scale structure (LSS), the stellar mass distribution throughout 3D space is not necessarily expected to be uniform, which is reflected in the noisy redshift evolution of the fit parameters, particularly $M^{*}$ and the two normalisation parameters. GAMA is overall underdense at $z<0.1$ \citep{driverGalaxyMassAssembly2022b}, which is also confirmed by the significantly lower $M^{*}$ and $\phi_{1}$ at the lowest redshift compared to the results at higher redshifts. The COSMOS field also contains a large number of galaxy clusters at $0.36<z<1$ \citep{finoguenovXMMNewtonWideFieldSurvey2007,bellagambaOptimalFilteringOptical2011}.

To mitigate LSS, each of the five double Schechter function parameters were fit with quadratic polynomials in redshift, as shown by the filled lines in each of the bottom panels in Figure~\ref{fig:smf_regressed}. This allowed us to compute a smooth, regressed SMF in each redshift bin, shown in the top right panel in Figure~\ref{fig:smf_regressed}. This was the same kind of fit employed by \citet{wrightGAMAG10COSMOS3DHST2018,thorneDeepExtragalacticVIsible2021} for their own calculations of the evolution of the SMF. The fit parameters of the double Schechter functions and the regressed fits are presented in Table~\ref{tab:smf_par} and Table~\ref{tab:smf_par_evol}. 

\subsection{LSS correction}\label{sec:lss}
Multiplying the $x-$ and $y-$axes of the SMFs in Figure~\ref{fig:smf_regressed} results in the stellar mass density distributions and their integrals give the CSMH as a function of redshift, i.e., 
\begin{equation}\label{eq:cdmh}
    \rho_{\mathrm{\star}}(\mathrm{z}) = \int^{\infty}_{0} \phi(\mathrm{M_{\star},z}) \, \times \, \mathrm{M_{\star}} \, \mathrm{dM_{\star}}.
\end{equation}

\begin{figure*}
    \centering
    \includegraphics[width=\textwidth]{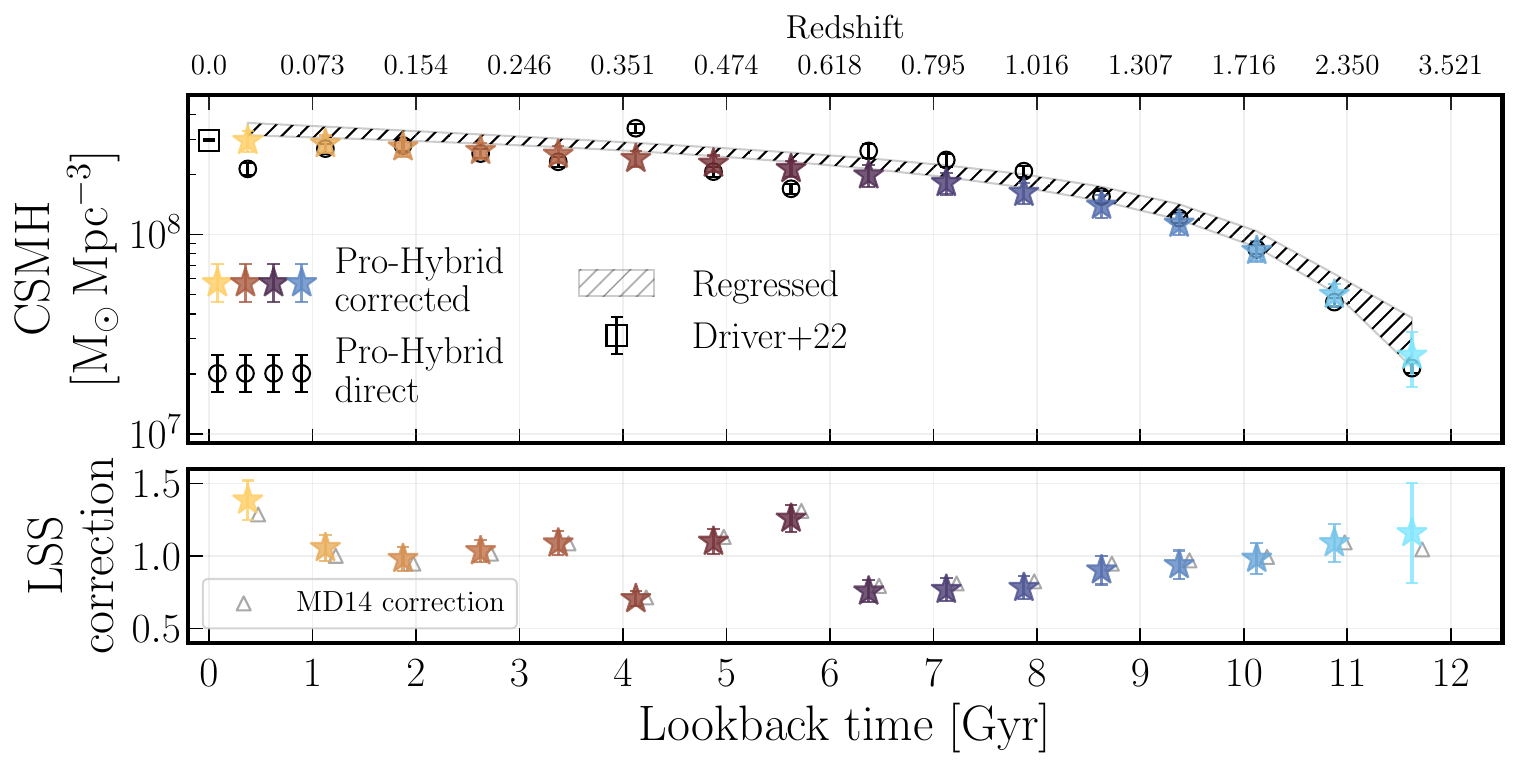}
    \caption{\textit{Top:} the CSMH. The black circles with error bars are the values and $1\sigma$ uncertainties from the direct double Schechter function fits to the SMF. The hatched grey line is the CSMH and $1\sigma$ uncertainties from the regressed SMFs. The coloured stars and error bars are the values and $1\sigma$ uncertainties corrected for LSS. The square and error bars is the measurement and $1\sigma$ uncertainty at $z-0$ from \citet{driverGalaxyMassAssembly2022b}. \textit{Bottom:} the LSS correction factor and $1\sigma$ uncertainties, obtained by comparing the smooth, corrected CSMH to the direct CSMH. \referee{The grey triangles show the expected LSS correction factor when using the integrated cosmic star formation history from \citet{madauCosmicStarFormationHistory2014} for the smooth CSMH. These have been offset from the star symbols for clarity.}}
    \label{fig:csmh}
\end{figure*}
Figure~\ref{fig:csmh} shows the CSMH. Signatures of LSS are evident in the noisy behaviour of the CSMH obtained from integrating the direct fits to the SMFs, particularly with our lowest redshift data point being $\approx 0.14$~dex lower than the $z=0$ measurement from \citet{driverGalaxyMassAssembly2022b} in the GAMA fields. \referee{The same noisiness due to LSS is seen in the evolution of the characteristic mass at the knee of the double Schechter, $10^{M^{*}}$, and the number density parameters, $\phi_{1}, \, \phi_{2}$. The motivation for fitting simple quadratic functions was to construct a data-driven smooth evolution of the CSMH. The normalisation of the smooth, regressed CSMH was then fit to the direct CSMH, previously obtained from the individually fitted SMFs, to compute a final smooth and corrected CSMH.} 

\referee{Dividing the direct CSMH with the corrected CSMH allowed us to compute a global LSS correction factor, shown in the bottom panel of Figure~\ref{fig:csmh}, that is the factor by which the stellar mass density at each redshift must be multiplied to account for the presence of LSS.} The corrected CSMH agrees well with \citet{driverGalaxyMassAssembly2022b} at $z=0$, validating this approach. The uncertainties on the LSS correction factors were propagated through to the uncertainties on the CSMH. 

\referee{Instead of fitting quadratic functions to the redshift evolution of the double Schechter functions, the smooth CSMH can also be calculated from the integration of the cosmic star formation history. For demonstration, we used the cosmic star formation history from \citet{madauCosmicStarFormationHistory2014} and an assumed return fraction of $R=0.41$ (appropriate for a \citet{chabrierGalacticStellarSubstellar2003} IMF) to compute an independent LSS correction factor for the CSMH, finding agreement with our data-driven results from the regressed CSMH.}

\section{Cosmic dust mass history}\label{sec:cdmh}
Having accounted for the stars in the cosmic baryon inventory in Section~\ref{sec:fitting_smf}, we now turn our attention to the dust. We computed the dust mass distribution functions (DMFs) using the same calculations as for the stars. 
\begin{figure*}
    \centering
    \includegraphics[width=\linewidth]{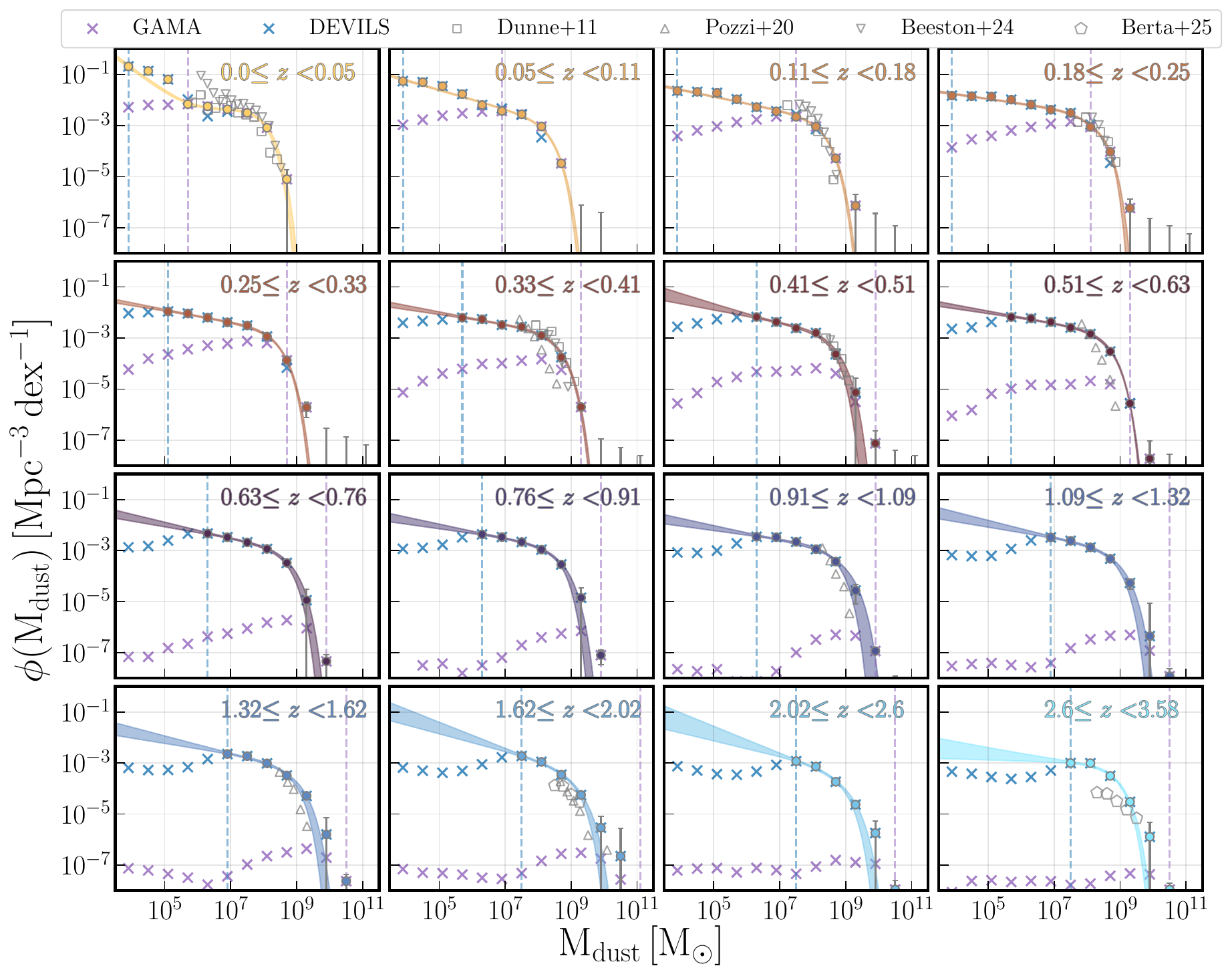}
    \caption{The DMF at $0<z\lesssim 3$ as indicated in the legend. In every panel, the solid points with $1\sigma$ error bars are the combined binned quantities from GAMA and DEVILS. The filled curves show the double Schechter functions and the $1\sigma$ uncertainty range. The fit parameters are presented in Table~\ref{tab:dmf_par}. The binned quantities for GAMA and DEVILS are shown independently with purple and blue crosses. \referee{Completeness limits for both GAMA and DEVILS are shown with dashed lines. Where the data points overlap between the surveys we prioritised the points with the lower uncertainty when fitting the DMFs.} We show results from \citet{dunneHerschelATLASRapidEvolution2011,pozziDustMassFunction2020,beestonConfirmingEvolutionDust2024,bertaPanchromaticViewN2CLS2025} with grey symbols as indicated in the legend.}
    \label{fig:dmf}
\end{figure*}
Figure~\ref{fig:dmf} shows the fitted DMFs for the $\mathrm{Pro-Hybrid}$ set in each of our 16 redshift bins. It can be seen that overall, the double Schechter functions adequately represent the data. Notably, the completeness limits that we derived from the maxima of the distributions of each GAMA and DEVILS indeed show that the method of combining these datasets in this way is valid. DEVILS dominates the contribution to the DMF at the low-mass/faint end at all redshifts and GAMA dominates the contribution at the massive end, similar to the SMFs. The DMFs as a function of redshift show very similar evolution as the SMFs, with a high-mass component becoming more prominent toward $z\approx 0$ \citep[e.g.,][]{baldryGalaxyMassAssembly2012,davidzonCOSMOS2015GalaxyStellar2017,grazianGalaxyStellarMass2015,lejaNewCensus022020,muzzinEVOLUTIONStelLARMASS2013,thorneDeepExtragalacticVIsible2021,weaverCOSMOS2020GalaxyStellar2023a,wrightGAMAG10COSMOS3DHST2018}.

We compare our DMFs to results from \citet{dunneHerschelATLASRapidEvolution2011,pozziDustMassFunction2020,beestonConfirmingEvolutionDust2024,bertaPanchromaticViewN2CLS2025} who all used SED fitting techniques to infer the dust masses. The fit parameters of the double Schechter functions presented in Table~\ref{tab:dmf_par}. There is overall agreement between our work and those from the literature. 

\begin{figure*}
    \centering
    \includegraphics[width=\linewidth]{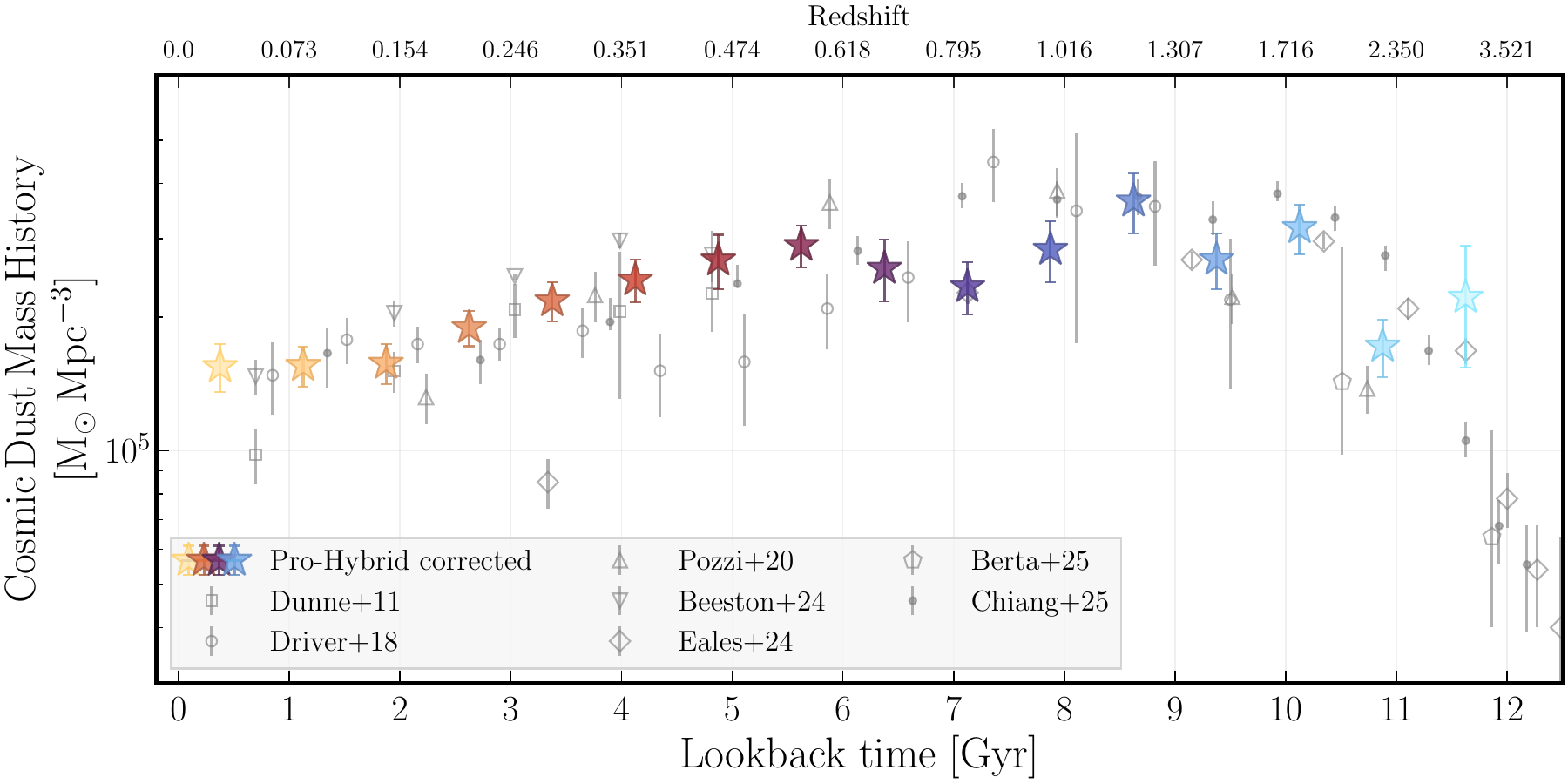}
    \caption{The CDMH. The coloured stars with error bars are the values and $1\sigma$ uncertainties using the $\mathrm{Pro-{Hybrid}}$ SED fits. 
    This has been corrected for LSS using the factors in Figure~\ref{fig:csmh}. We show previous measurements and $1\sigma$ uncertainties as indicated \citep{dunneHerschelATLASRapidEvolution2011,driverGAMAG10COSMOS3DHST2018a,pozziDustMassFunction2020,beestonConfirmingEvolutionDust2024,ealesRiseFallDust2024,chiangCosmicInfraredBackground2025a, bertaPanchromaticViewN2CLS2025}.}
    \label{fig:cdmh}
\end{figure*}
Figure~\ref{fig:cdmh} shows the CDMH computed from the fitted DMFs. Some tension is noticeable at $0.35 \lesssim z \lesssim 0.6$ where our results are $\approx 0.2$~dex lower than, for example, \citet{driverGAMAG10COSMOS3DHST2018a}. However, this may be due to cosmic variance because, as can be seen from the LSS correction factors in Figure~\ref{fig:csmh}, there is some indication that the COSMOS region is under-dense compared to the cosmic mean in that redshift range \citep{bellagambaOptimalFilteringOptical2011} and could have skewed their results to lower values. \referee{Our results are also $\approx 0.2$~dex lower than \cite{driverGAMAG10COSMOS3DHST2018a} at $z\approx 0.8$ that is also likely due to our LSS correction.}

\referee{Over all, our results on the CDMH are consistent with previous measurements from the literature across $0<z<3$ \citep{dunneHerschelATLASRapidEvolution2011,driverGAMAG10COSMOS3DHST2018a,pozziDustMassFunction2020,beestonConfirmingEvolutionDust2024,ealesRiseFallDust2024,bertaPanchromaticViewN2CLS2025,chiangCosmicInfraredBackground2025a}. The CDMH peaks at $z\approx 1.5$, having increased by a factor of $\approx 1.8$ from $z\approx3$. By $z\approx 0$ the CDMH has declined by a factor of $z\approx 2.2$ from its maximum.} 

It is worth discussing the metallicity-dependent DTH correction, as described in Section~\ref{sec:modellingAssumptions}. \referee{The slope of the CDMH at $z \lesssim 1$ is similar when using $\mathrm{DTH=0.0073}$ versus the \citet{remy-ruyerGastodustMassRatios2014a} metallicity-dependent DTH. The reason for this is that the Universe is at this point on average enriched to at least solar metallicity. At higher redshift, the CDMH inferred with the solar metallicity DTH was basically flat, meaning that the metallicity-dependent correction was important to better resolve the peak of the CDMH, as opposed to a plateau, at $z \gtrsim 1$.}

\subsection{CDMH Comparison to simulations}
The corrected CDMH generally follows a similar evolution to the cosmic star formation history with a \referee{peak $z\approx 1.5$}. It can be seen that the slope of the CDMH below $z\approx 0.5$ is shallower than the slope of the cosmic star formation history \citep{dsilvaGAMADEVILSCosmic2023,dsilvaSelfConsistentJWSTCensus2025a}. \referee{The decreasing accretion rate of matter onto dark matter haloes is believed to have driven the decline of the cosmic star formation history over the last $10$~Gyr \citep[e.g.,][]{kauffmannUnifiedModelEvolution2000, schayePhysicsDrivingCosmic2010b, madauCosmicStarFormationHistory2014, behrooziUniverseMachineCorrelationGalaxy2019,harikaneGOLDRUSHIVLuminosity2022a}. Furthermore, feedback from stars and AGN in galaxies likely decreases the efficiency at which H$_{2}$ is converted from HI, resulting in a flattening of the atomic gas mass density \citep[e.g.,][]{walterEvolutionBaryonsAssociated2020,wrightImpactStellarAGN2020,wrightBaryonCycleModern2024b}. The down turn in the population-averaged star formation rate then drives the decline of the CDMH.}

Dust is formed from metals that are themselves produced during stellar evolution \citep[e.g.,][]{draineInterstellarDustGrains2003,ginolfiWhereDoesGalactic2018} through the mechanisms of stellar winds off AGB stars \citep{venturaDustAsymptoticGiant2014,hofnerMassLossStars2018,hofnerExplainingWindsAGB2020} and supernovae explosions \citep{nomotoNucleosynthesisYieldsCorecollapse2006,bocchioDustGrainsHeart2016,marassiSupernovaDustYields2019,sarangiDustSupernovaeSupernova2018,schneiderFormationCosmicEvolution2024}, and so the expectation is that the CDMH is \referee{connected to} the cosmic star formation history. At the same time, dust grains in the ISM may be destroyed or consumed via star formation \citep{draineDestructionMechanismsInterstellar1979,drainePhysicsDustGrains1979,jonesGrainDestructionShocks1994,jonesGrainShatteringShocks1996,micelottaDustSupernovaeSupernova2018}. \referee{Dust grains may also grow in the ISM through accretion, which dominates the contribution to the dust mass density, and is most efficient in dense molecular clouds and appears to dominate the contribution to the dust mass density \citep[e.g.,][]{savageInterstellarAbundancesAbsorptionLine1996,hirashitaDustGrowthTimescale2000,hirashitaDustGrowthInterstellar2012,asanoDustFormationHistory2013,trayfordModellingEvolutionInfluence2026}.} The CDMH is hence a useful quantity to compare against simulations as it depends intricately on the sub-grid implementations of feedback, star formation \referee{and dust growth}.

\begin{figure}
    \centering
    \includegraphics[width=\columnwidth]{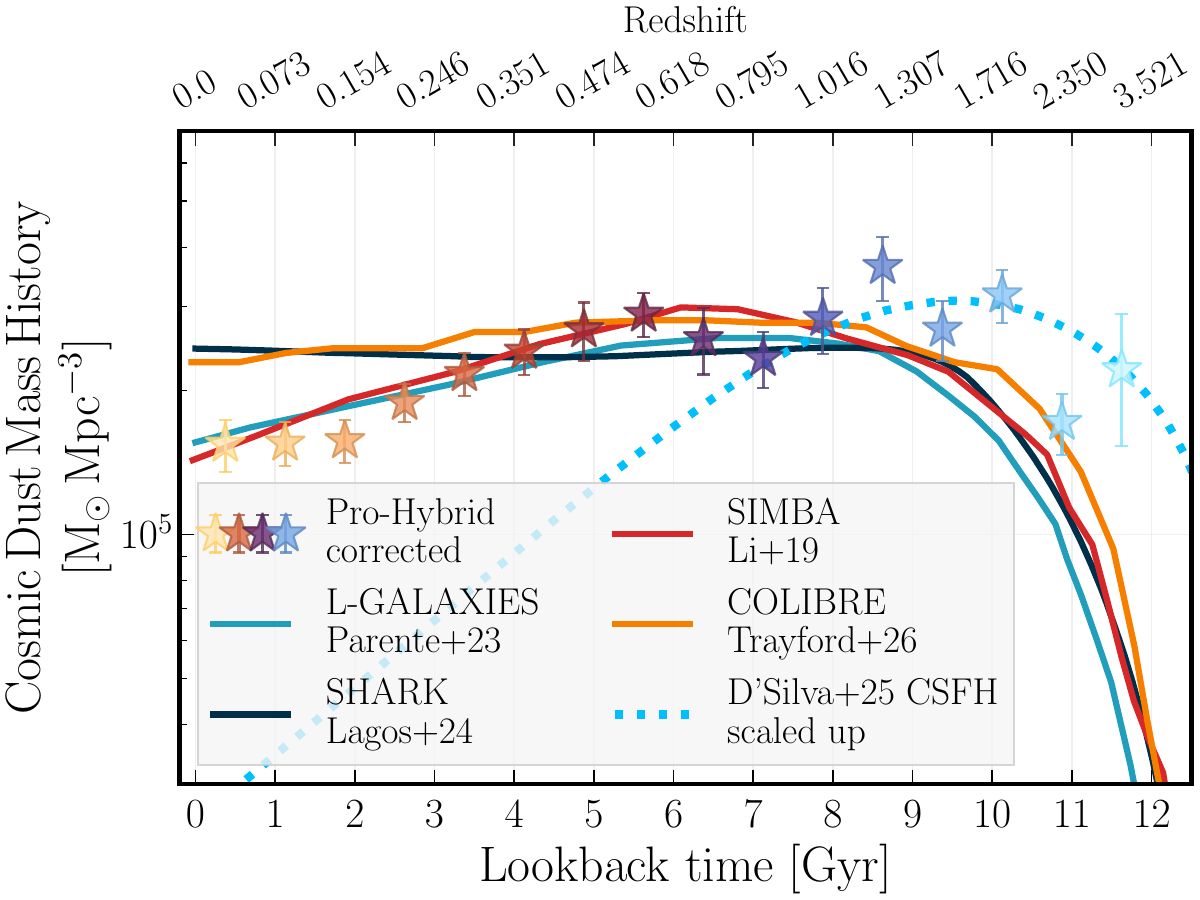}
    \caption{Comparison of the CDMH with simulations. The colourful points and error bars are the same as in Figure~\ref{fig:cdmh}. We show simulation predictions \citep{liDusttogasDusttometalRatio2019,parente1DropCosmic2023,lagosQuenchingMassiveGalaxies2024,trayfordModellingEvolutionInfluence2026} as indicated in the figure legend. The dotted blue line is the cosmic star formation history from \citet{dsilvaGAMADEVILSCosmic2023,dsilvaSelfConsistentJWSTCensus2025a} scaled up by $\approx 4\times10^{6}$.}
    \label{fig:cdmh_sims}
\end{figure}

Indeed, the consensus from simulations is that the decline of the CDMH closely follows the cosmic star formation history \citep{poppingDustContentGalaxies2017,vijayanDetailedDustModelling2019,trianiOriginDustGalaxies2020,parente1DropCosmic2023}. \referee{The dust destruction and astration rates decline commensurately with the decline of star formation, meaning that dust grains may survive more easily in the ISM  at $z\lesssim 1-2$, past the peak of cosmic star formation density. At the same time, though the volume-averaged star formation rate declines, dust may still be formed by AGB stars and when stars go supernova. Both of these factors combined indicate that the CDMH declines with the cosmic star formation history past the peak, though with a shallower slope \citep[e.g.,][]{yatesImpactBinaryStars2024,osmanDustEvolutionCosmic2025}.} In Figure~\ref{fig:cdmh_sims} we compare our CDMH with semi-analytic models: L-\textsc{Galaxies} \citep{parente1DropCosmic2023} and \textsc{Shark} \citep{lagosQuenchingMassiveGalaxies2024}, and hydrodynamic models: \textsc{SIMBA} \citep{daveSIMBACosmologicalSimulations2019a,liDusttogasDusttometalRatio2019} and COLIBRE \citep{schayeCOLIBREProjectCosmological2025,trayfordModellingEvolutionInfluence2026}.

In general, our fully corrected CDMH is mostly in agreement with these simulations at $z \lesssim $\referee{1}. We remark that \textsc{Shark} and COLIBRE predict a shallower decline below $z\approx 0.4$ than both L-GALAXIES and SIMBA. \referee{At $z<0.4$ we more closely agree with L-GALAXIES and SIMBA, whereas we find $\approx 0.2$~dex lower values compared to \textsc{Shark} and COLIBRE.} The differences between these two sets of simulations at $z\lesssim 0.4$ are more than likely due to their implementation of stellar feedback that both suppresses the star formation/chemical enrichment and destroys the dust/expels it from the galaxy. \referee{At $z>1$, our measurements of the CDMH are on average $\approx 0.2-0.6$~dex higher than all simulations considered.}

\section{The baryon inventory}\label{sec:baryonInventory}
\begin{figure}
    \centering
    \includegraphics[width=\columnwidth]{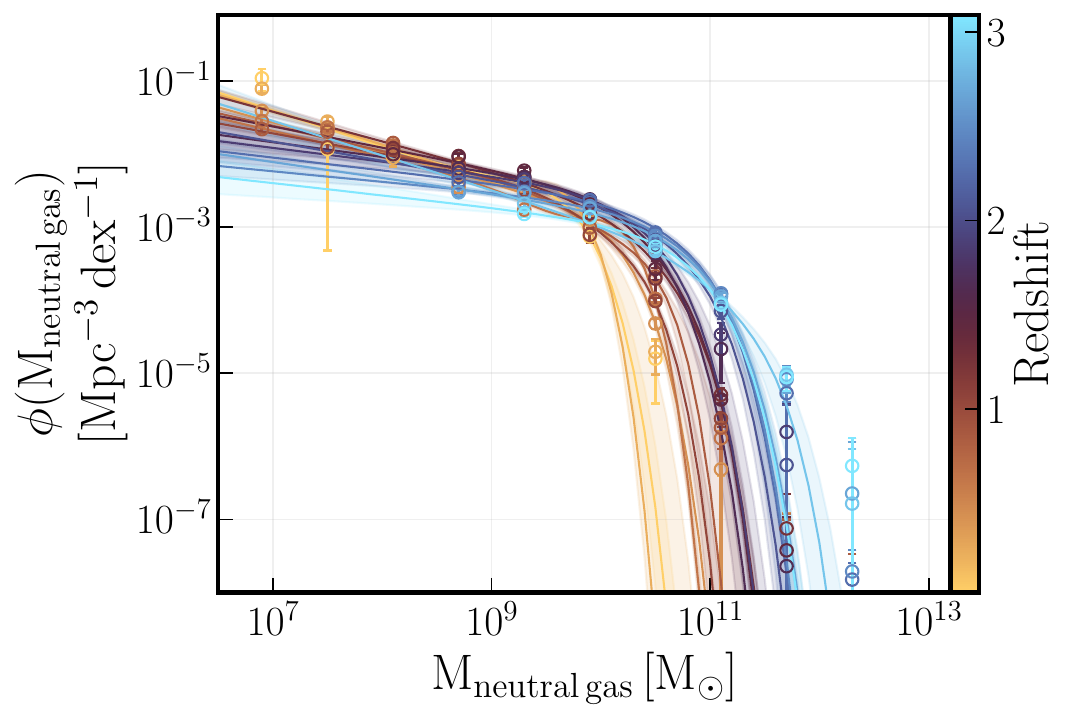}
    \caption{The dust-traced GMFs at $0<z\lesssim 3$, as indicated in the legend, for $\mathrm{Pro-Hybrid}$. The filled curves show the double Schechter functions and the $1\sigma$ uncertainty range. The fit parameters are presented in Table~\ref{tab:gmf_par}.}
    \label{fig:gmf}
\end{figure}
In addition to the DTH, \citet{remy-ruyerGastodustMassRatios2014a} also report the dust-to-gas mass ratio \citep[DTG, see also][]{deciaDustdepletionSequencesDamped2016,wisemanEvolutionDusttometalsRatio2017,devisSystematicMetallicityStudy2019,heintzCosmicBuildupDust2023,yatesImpactBinaryStars2024} \referee{that is ratio between the dust mass and the combined mass in hydrogen (HI and H$_{2}$), helium and gaseous metals. In detail, 
\begin{equation}
    \mathrm{M_{gas} = \mu_{gal}(M_{HI} + M_{H_{2}})},
\end{equation} where $\mathrm{\mu_{gal}} = 1 / (1-Y_{\odot}-Z)$ is the mean atomic weight, $Y_{\odot} = 0.270$ is the Galactic mass fraction of helium \citep{asplundChemicalCompositionSun2009a} and $Z$ is the metallicity.
}

The metallicity-dependent DTG was used to convert the GAMA and DEVILS dust masses to total neutral gas masses, using the $\mathrm{Z_{final}}$ parameter from \textsc{ProSpect}. With the neutral gas masses we computed the neutral gas mass distribution functions (GMFs) that were fitted with double Schechter functions. The results are shown in Figure~\ref{fig:gmf} and the fit parameters are presented in Table~\ref{tab:gmf_par}. 

\begin{figure*}
    \centering
    \includegraphics[width=\linewidth]{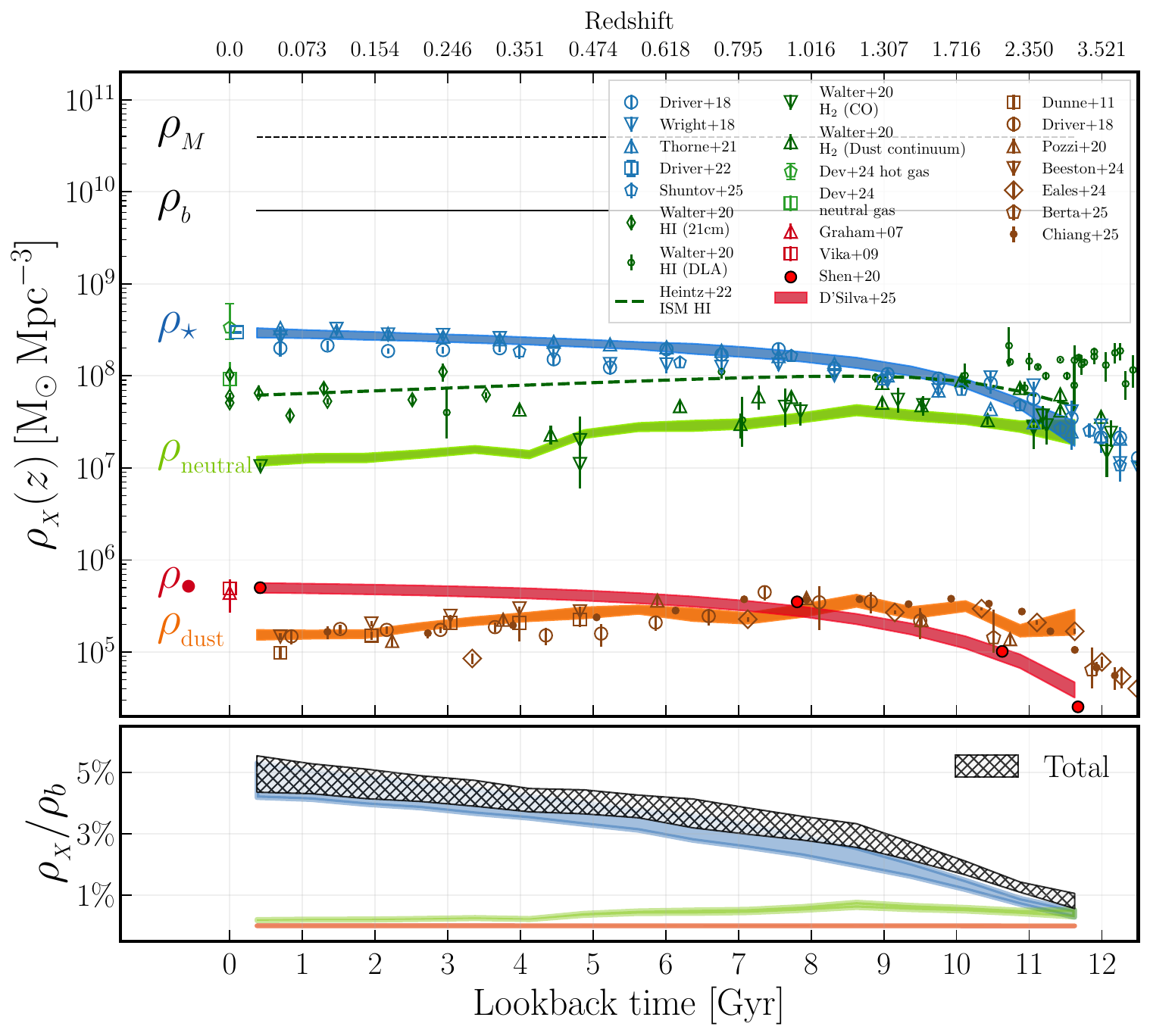}
    \caption{\textit{Top:} cosmic mass densities and the baryon inventory. The blue curve shows the CSMH from the regressed fits of the SMFs. The orange filled curve shows the results of the CDMH. The green filled curve is the CGMH. All of these have had the LSS correction applied. The red curve shows the supermassive black hole mass history from \citet{dsilvaGAMADEVILSCosmic2023,dsilvaSelfConsistentJWSTCensus2025a}, which is inferred from the AGN luminosity history by assuming a 10 per cent radiative efficiency and successively integrating down to $z=0$. The solid line is the baryon density and the dashed line is the total matter density from $\Lambda$CDM. Observations and $1\sigma$ uncertainties of the stellar mass densities \citep{driverGAMAG10COSMOS3DHST2018a,wrightGAMAG10COSMOS3DHST2018,thorneDeepExtragalacticVIsible2021,driverGalaxyMassAssembly2022b,shuntovCOSMOSWebStellarMass2025}, gas mass densities \citep{walterEvolutionBaryonsAssociated2020,devBaryonCensusMassdensity2024a}, SMBH mass densities \citep{grahamMillenniumGalaxyCatalogue2007,vikaMillenniumGalaxyCatalogue2009,shenBolometricQuasarLuminosity2020} and the same literature results of the dust mass densities from Figure~\ref{fig:cdmh} are shown in blue, green, red and brown as indicated in the figure legend. \textit{Bottom:} baryon fraction obtained by dividing the curves in the top panel with the baron density. The total baryon fraction in galaxies is shown as the grey filled region. The colour scheme is the same as the top panel. The results for the stars, dust and neutral gas are presented in Table~\ref{tab:rho_baryon}.}
    \label{fig:rho}
\end{figure*}
Integrating the GMFs we computed the cosmic neutral gas mass history (CGMH), shown as the green band in Figure~\ref{fig:rho}. 

The shape of the CGMH inferred from the dust mass shows remarkable agreement with previous results from the compilations presented in \citet{walterEvolutionBaryonsAssociated2020}. It is however clear that the normalisation of our CGMH underestimates the HI mass density, especially at $z<1$ where the measurements are from $21$~cm observations \citep[though there is scatter between the results from the compilation of][]{walterEvolutionBaryonsAssociated2020}. \referee{The HI compilation from \citet[][see their Tab. 3 and associated references]{walterEvolutionBaryonsAssociated2020} includes measurements of the HI mass density obtained from direct $21$~cm observations at $z = 0$, stacked $21$~cm observations at $z= 0-1$, $21$~cm cross-correlation results from $z = 0.2-1.8$ and damped Lyman alpha (DLA) systems at $z\approx 0-5$. The average difference over all redshift is $\approx 0.7$~dex, with our points being lower than those from \citet{walterEvolutionBaryonsAssociated2020}. We also show the results from \citet{heintzALMAREBELSSurvey2022} that is the HI mass in galaxies, obtained by scaling our cosmic stellar mass history with their HI to stellar mass ratio as a function of redshift, $\mathrm{M_{HI}/M_{\star}} \propto (1+z)^{1.6}$. Compared to \citet{heintzALMAREBELSSurvey2022}, our measurements are $\approx 0.5$~dex lower at $z\approx 3$ and $\approx 0.7$~dex lower at $z\approx 0$.}

We expect that our lower CGMH is due to differences in the spatial distributions of dust, from which our neutral gas masses are inferred, and HI in galaxies. Our dust masses are confined within the optical radius as per the source detection, photometric extraction and SED fitting. In contrast, the mass of HI within the optical radius can be \referee{as low as $20$~per cent} of the total HI mass \citep[e.g.,][]{leeWALLABYPilotSurvey2025}. Moreover, our lowest redshift point for the neutral gas density is \referee{$\approx 0.9$~dex} lower than \citet{devBaryonCensusMassdensity2024a}, potentially because \citet{devBaryonCensusMassdensity2024a} calculated the gas content in the dark matter haloes, covering a larger spatial extent of neutral gas than traced by dust. The CGMH tends to more closely agree with the $\mathrm{H_{2}}$ gas mass density from the compilation in \citet{walterEvolutionBaryonsAssociated2020}, where, unlike the HI, the molecular gas and dust are expected to have similar spatial distributions in galaxies. We note that the DTG ratio that we used was calculated within the aperture of the dust emission \citep{remy-ruyerGastodustMassRatios2014a} and so does not account for neutral gas beyond the optical radius.

The contribution to the baryon inventory from SMBHs was inferred from the cosmic AGN luminosity history as presented in \citet{dsilvaGAMADEVILSCosmic2023,dsilvaSelfConsistentJWSTCensus2025a} who used the same GAMA and DEVILS data also used here to compute this quantity at $0<z<3$. The cosmic SMBH history was inferred by assuming a $10$ per cent radiative efficiency and successively integrating down to $z=0$ \citep[e.g.,][]{shakuraBlackHolesBinary1973}.

In Figure~\ref{fig:rho}, we compare all of our baryon measurements to a swath of results from the literature of the cosmic stellar mass densities \citep{driverGAMAG10COSMOS3DHST2018a,wrightGAMAG10COSMOS3DHST2018,thorneDeepExtragalacticVIsible2021,driverGalaxyMassAssembly2022b,shuntovCOSMOSWebStellarMass2025}, neutral gas mass densities \citep{walterEvolutionBaryonsAssociated2020,devBaryonCensusMassdensity2024a}, SMBH mass densities \citep{grahamMillenniumGalaxyCatalogue2007,vikaMillenniumGalaxyCatalogue2009,shenBolometricQuasarLuminosity2020} and, as previously discussed, dust mass densities \citep{dunneHerschelATLASRapidEvolution2011,driverGAMAG10COSMOS3DHST2018a,pozziDustMassFunction2020,beestonConfirmingEvolutionDust2024,ealesRiseFallDust2024,chiangCosmicInfraredBackground2025a, bertaPanchromaticViewN2CLS2025}. We find overall agreement between our baryon census and previous results from the literature. The novelty of our work is demonstrated in Figure~\ref{fig:rho} because every quantity of the mass density was calculated using essentially the same data (GAMA and DEVILS) and methods (i.e., \textsc{ProSpect}) throughout, meaning that this census of the baryons is self-consistent. The results for the stars, dust and neutral gas are presented in Table~\ref{tab:rho_baryon}.

The baryon inventory of today is quantified in relation to the cosmic baryon density, which as predicted by $\Lambda$CDM is $\Omega_{b} \approx 0.05$ \citep{aghanimPlanck2018Results2020c}, in units of the critical density, $\rho_{\mathrm{crit,}0} = \mathrm{\frac{3\,H_{0}^{2}}{8\pi \, G}}$. We find that stars, neutral gas, SMBHs and dust within the optical radii of galaxies account for $\approx 5$ per cent of the baryon budget. This value has grown from $\approx 1$ per cent over the last $\approx 12$~Gyr up to the present. Notably, we have not accounted for hot gas in the interstellar, circumgalactic or intergalactic medium \citep[e.g.][]{ferriereInterstellarEnvironmentOur2001,drainePhysicsInterstellarIntergalactic2011,tumlinsonCircumgalacticMedium2017}. The measurement of the hot, X-ray emitting gas density from \citet{devBaryonCensusMassdensity2024a} is at the same level as the stellar mass at $z=0$. Hence, many of the remaining baryons are expected to be spread throughout the dark matter haloes of galaxies and the cosmic web \citep[e.g.,][]{fukugitaCosmicEnergyInventory2004,graaffProbingMissingBaryons2019} in the form of ionised gas \citep{macquartCensusBaryonsUniverse2020,wrightBaryonCycleModern2024b,connorGasrichCosmicWeb2025}

\section{Conclusions}\label{sec:conclusion}
In this paper, we have investigated the volume-averaged evolution of stars, neutral gas, SMBHs and dust in galaxies over the last $\approx 12$~Gyr using GAMA and DEVILS. The main results are summarised below:

\begin{enumerate}
    \item We computed the stellar, dust and neutral gas mass distribtion functions in 16 redshift bins, uniformly spaced in $0.75$~Gyr intervals of lookback time. Each of these were fitted with double Schechter functions.
    
    \item Integrating those we obtained the cosmic stellar, dust and neutral gas mass history at $0<z<3$. The cosmic dust mass history in particular follows a similar shape to the cosmic star formation history, peaking at $z \approx 1-2$ and declining thereafter up to the present. The decline is interpreted as a balance between the slowing down of dust production and destruction, both due to the decrease of star formation down to $z=0$.
    
    \item Using the dust masses directly from \textsc{ProSpect} resulted in $\approx 3$ times higher dust masses than previous results with the SED fitting code MAGPHYS. In fact, this difference will persist between any code that also employs modified blackbody dust emission. \referee{Part of the} reason behind this is due to the assumption of a uniform mass-to-light ratio assumed for all IR-FIR emitting dust species in \textsc{ProSpect}. If we instead assumed that very small grains and PAH molecules, which emit predominantly at $\lambda \lesssim 2\times10^{5} \SI{}{\angstrom}$, contribute no more than \referee{$\approx 1-3.5$ per cent} of the total dust mass, the resulting dust masses came into better agreement with MAGPHYS. \referee{Furthermore, the use of a metallicity-dependent DTH ratio, as opposed to the DTH ratio calculated at solar metallicity, was important for inferring dust masses with \textsc{ProSpect}.}

    \item We compared the CDMH with both the semi-analytic models, \textsc{Shark} and L-GALAXIES, and the hydrodynamic models, SIMBA and COLIBRE. We found that all of the simulations are mostly in agreement with the CDMH at \referee{$z\lesssim 1$}. \textsc{Shark} and COLIBRE exhibit a shallower decline at $z<0.4$ compared to SIMBA and L-GALAXIES pointing to differences in the implementation of dust growth and destruction between the two sets of models. We estimated \referee{$\approx 0.2-0.6$~dex} higher values of the CDMH at $z>1$ compared to all simulations.

    \item Dividing the dust masses by the metallicity-dependent dust-to-gas ratio of \citet{remy-ruyerGastodustMassRatios2014a}, we obtained neutral gas masses from which we calculated the neutral gas mass distribution functions and the cosmic neutral gas mass density, finding broad agreement with previous results. The gas mass density traced by the dust is \referee{$\approx 0.7$~dex} lower than that inferred from $21$cm observations, more than likely because of the different spatial extents covered by dust and HI in galaxies.
    
    \item Folding in previous measurements of the cosmic SMBH mass densities that were also computed on the same data sets of GAMA/DEVILS and in an identical manner, we presented a self-consistent and homogeneous census of the baryon inventory in galaxies. Stars, neutral gas, SMBHs and dust within the optical radii of galaxies constitute about $5$ per cent of the baryon budget. This means that most of the remaining $95$ per cent of baryons must be diffuse ionised gas residing in the interstellar, circumgalactic and intergalactic medium.
\end{enumerate}

\section*{Acknowledgements}
\referee{We thank the anonymous referee for their constructive feedback.} JCJD acknowledges support from the Australian Government Research Training Program (RTP) Scholarship. CL is a recipient of the ARC Discovery Project DP210101945.  AJB. and EdC. acknowledge support from the Australian Research Council (project DP240100589).

DEVILS is an Australian project based around a spectroscopic campaign using the Anglo-Australian Telescope. DEVILS is part funded via Discovery Programs by the Australian Research Council and the participating institutions. The DEVILS website is \url{devils.research.org.au}. 

\section*{Data Availability}
The analysis scripts and data can be found on \textsc{GitHub} here: \url{https://github.com/JordanDSilva/dust-mass-density}. Data from the GAMA survey may be found here: \url{https://www.gama-survey.org}. The DEVILS data are hosted and provided by AAO Data Central (\url{https://datacentral.org.au}).


\bibliographystyle{mnras}
\bibliography{ref} 




\appendix

\section{Fit parameters of distribution functions}
\begin{table*}
    \centering
    \begin{tabular}{cccccc}
    $\left< z \right>$ & $\mathrm{\log_{10}(M^{*} \, [M_{\odot}]})$ & $\mathrm{\alpha}$ & $\mathrm{\beta}$ & $\mathrm{\log_{10}(\phi_{1} \, [Mpc^{-3}])}$ & $\mathrm{\log_{10}(\phi_{2} \, [Mpc^{-3}])}$ \\
    \hline
    $0.027$ & $10.742 \pm 0.055$ & $-1.766 \pm 0.024$ & $-0.556 \pm 0.084$ & $-3.776 \pm 0.091$ & $-2.420 \pm 0.061$ \\
    $0.084$ & $10.983 \pm 0.026$ & $-1.658 \pm 0.030$ & $-0.567 \pm 0.087$ & $-3.364 \pm 0.075$ & $-2.696 \pm 0.059$ \\
    $0.146$ & $11.021 \pm 0.026$ & $-1.563 \pm 0.026$ & $-0.531 \pm 0.086$ & $-3.302 \pm 0.069$ & $-2.714 \pm 0.067$ \\
    $0.214$ & $11.049 \pm 0.032$ & $-1.460 \pm 0.020$ & $-0.467 \pm 0.095$ & $-3.098 \pm 0.054$ & $-2.955 \pm 0.119$ \\
    $0.288$ & $11.126 \pm 0.044$ & $-1.495 \pm 0.024$ & $-0.519 \pm 0.095$ & $-3.202 \pm 0.055$ & $-3.135 \pm 0.166$ \\
    $0.371$ & $11.044 \pm 0.029$ & $-1.383 \pm 0.027$ & $-0.513 \pm 0.097$ & $-2.874 \pm 0.060$ & $-2.887 \pm 0.135$ \\
    $0.464$ & $11.101 \pm 0.034$ & $-1.401 \pm 0.027$ & $-0.516 \pm 0.094$ & $-3.126 \pm 0.062$ & $-3.224 \pm 0.163$ \\
    $0.569$ & $11.001 \pm 0.024$ & $-1.471 \pm 0.054$ & $-0.535 \pm 0.096$ & $-3.248 \pm 0.099$ & $-3.050 \pm 0.118$ \\
    $0.690$ & $10.969 \pm 0.074$ & $-1.445 \pm 0.057$ & $-0.489 \pm 0.102$ & $-3.159 \pm 0.116$ & $-2.701 \pm 0.116$ \\
    $0.832$ & $10.980 \pm 0.049$ & $-1.439 \pm 0.053$ & $-0.491 \pm 0.096$ & $-3.220 \pm 0.102$ & $-2.758 \pm 0.099$ \\
    $1.001$ & $11.038 \pm 0.055$ & $-1.356 \pm 0.062$ & $-0.522 \pm 0.102$ & $-3.185 \pm 0.123$ & $-2.947 \pm 0.143$ \\
    $1.208$ & $11.014 \pm 0.058$ & $-1.485 \pm 0.074$ & $-0.492 \pm 0.093$ & $-3.399 \pm 0.116$ & $-3.020 \pm 0.129$ \\
    $1.470$ & $11.025 \pm 0.039$ & $-1.492 \pm 0.076$ & $-0.512 \pm 0.092$ & $-3.502 \pm 0.119$ & $-3.166 \pm 0.107$ \\
    $1.818$ & $11.044 \pm 0.036$ & $-1.488 \pm 0.059$ & $-0.512 \pm 0.100$ & $-3.548 \pm 0.087$ & $-3.484 \pm 0.139$ \\
    $2.311$ & $11.108 \pm 0.038$ & $-1.535 \pm 0.046$ & $-0.492 \pm 0.096$ & $-3.837 \pm 0.076$ & $-3.995 \pm 0.152$ \\
    $3.090$ & $10.581 \pm 0.055$ & $-1.508 \pm 0.081$ & $-0.532 \pm 0.102$ & $-3.728 \pm 0.111$ & $-3.562 \pm 0.153$ \\
    \hline
    \end{tabular}
    \caption{Double Schechter function fit parameters and $1\sigma$ uncertainties for the SMFs. These fitted functions should be multiplied by the LSS correction factor to get the same CSMH as in Figure~\ref{fig:csmh} and Table~\ref{tab:rho_baryon}.}
    \label{tab:smf_par}
\end{table*}

\begin{table*}
    \centering
    \begin{tabular}{cccc}
    & $C_{0}$ & $C_{1}$ & $C_{2}$ \\
    \hline
    $\mathrm{\log_{10}(M^{*} \, [M_{\odot}]})$ & $10.963 \pm 0.017$ & $0.211 \pm 0.040$ & $-0.094 \pm 0.015$ \\
    $\alpha$ & $-1.627 \pm 0.015$ & $0.428 \pm 0.047$ & $-0.154 \pm 0.018$ \\
    $\beta$ & $-0.539 \pm 0.044$ & $0.053 \pm 0.090$ & $-0.017 \pm 0.031$ \\
    $\mathrm{\log_{10}(\phi_{1} \, [Mpc^{-3}])}$ & $-3.212 \pm 0.036$ & $0.059 \pm 0.091$ & $-0.105 \pm 0.033$ \\
    $\mathrm{\log_{10}(\phi_{2} \, [Mpc^{-3}])}$ & $-2.592 \pm 0.038$ & $-0.515 \pm 0.098$ & $0.045 \pm 0.038$ \\
    \hline
    \end{tabular}
    \caption{Regressed fits and $1\sigma$ uncertainties to each of the double Schechter function parameters for the SMFs. Fits are quadratic in redshift, $C_{0} + zC_{1} + z^{2}C_{2}$.}
    \label{tab:smf_par_evol}
\end{table*}

\begin{table*}
    \centering
    \begin{tabular}{cccccc}
    $\left< z \right>$ & $\mathrm{\log_{10}(M^{*} \, [M_{\odot}]})$ & $\mathrm{\alpha}$ & $\mathrm{\beta}$ & $\mathrm{\log_{10}(\phi_{1} \, [Mpc^{-3}])}$ & $\mathrm{\log_{10}(\phi_{2} \, [Mpc^{-3}])}$ \\
    \hline
    $0.027$ & $7.809 \pm 0.076$ & $-1.972 \pm 0.058$ & $-0.922 \pm 0.096$ & $-4.960 \pm 0.278$ & $-2.773 \pm 0.085$ \\
    $0.084$ & $8.152 \pm 0.063$ & $-1.376 \pm 0.019$ & $-0.372 \pm 0.853$ & $-3.169 \pm 0.056$ & $-5.240 \pm 1.866$ \\
    $0.146$ & $8.180 \pm 0.039$ & $-1.269 \pm 0.015$ & $-0.060 \pm 0.960$ & $-3.076 \pm 0.050$ & $-6.103 \pm 1.309$ \\
    $0.214$ & $7.988 \pm 0.102$ & $-1.183 \pm 0.017$ & $1.420  \pm 0.288$ & $-2.880 \pm 0.067$ & $-3.933 \pm 0.328$ \\
    $0.288$ & $8.291 \pm 0.038$ & $-1.236 \pm 0.025$ & $-0.060 \pm 1.023$ & $-3.079 \pm 0.066$ & $-5.930 \pm 1.397$ \\
    $0.371$ & $8.461 \pm 0.066$ & $-1.223 \pm 0.031$ & $0.305  \pm 1.048$ & $-3.002 \pm 0.070$ & $-5.270 \pm 1.479$ \\
    $0.464$ & $8.449 \pm 0.152$ & $-1.309 \pm 0.092$ & $-0.225 \pm 1.035$ & $-3.259 \pm 0.215$ & $-5.328 \pm 1.997$ \\
    $0.569$ & $8.502 \pm 0.085$ & $-1.217 \pm 0.029$ & $0.377  \pm 1.114$ & $-3.211 \pm 0.062$ & $-5.666 \pm 1.508$ \\
    $0.690$ & $8.585 \pm 0.118$ & $-1.262 \pm 0.048$ & $-0.054 \pm 0.971$ & $-3.172 \pm 0.109$ & $-5.484 \pm 1.680$ \\
    $0.832$ & $8.490 \pm 0.169$ & $-1.229 \pm 0.050$ & $0.345  \pm 1.012$ & $-3.102 \pm 0.123$ & $-4.944 \pm 1.722$ \\
    $1.001$ & $8.639 \pm 0.196$ & $-1.219 \pm 0.055$ & $0.384  \pm 1.055$ & $-3.186 \pm 0.160$ & $-5.297 \pm 1.528$ \\
    $1.208$ & $8.793 \pm 0.136$ & $-1.276 \pm 0.057$ & $0.055  \pm 1.013$ & $-3.312 \pm 0.124$ & $-5.449 \pm 1.603$ \\
    $1.470$ & $8.808 \pm 0.169$ & $-1.283 \pm 0.067$ & $0.450  \pm 0.996$ & $-3.486 \pm 0.156$ & $-5.371 \pm 1.449$ \\
    $1.818$ & $8.979 \pm 0.169$ & $-1.422 \pm 0.082$ & $0.293  \pm 0.977$ & $-3.675 \pm 0.168$ & $-5.696 \pm 1.495$ \\
    $2.311$ & $8.868 \pm 0.185$ & $-1.439 \pm 0.109$ & $0.283  \pm 1.021$ & $-3.897 \pm 0.216$ & $-5.830 \pm 1.290$ \\
    $3.090$ & $8.769 \pm 0.085$ & $-1.126 \pm 0.116$ & $-0.158 \pm 1.115$ & $-3.559 \pm 0.269$ & $-5.575 \pm 1.769$ \\
    \hline
    \end{tabular}
    \caption{Double Schechter function fit parameters and $1\sigma$ uncertainties for the DMFs. These fitted functions should be multiplied by the LSS correction factor to get the same CDMH as in Figure~\ref{fig:rho} and Table~\ref{tab:rho_baryon}.}
    \label{tab:dmf_par}
\end{table*}

\begin{table*}
    \centering
    \begin{tabular}{cccccc}
    $\left< z \right>$ & $\mathrm{\log_{10}(M^{*} \, [M_{\odot}]})$ & $\mathrm{\alpha}$ & $\mathrm{\beta}$ & $\mathrm{\log_{10}(\phi_{1} \, [Mpc^{-3}])}$ & $\mathrm{\log_{10}(\phi_{2} \, [Mpc^{-3}])}$ \\
    \hline
    $0.027$ & $9.398  \pm 0.124$ & $-1.499 \pm 0.019$ & $-0.275 \pm 0.418$ & $-3.119 \pm 0.120$ & $-2.662 \pm 0.128$ \\
    $0.084$ & $9.181  \pm 0.279$ & $-1.440 \pm 0.022$ & $1.072  \pm 0.851$ & $-2.759 \pm 0.158$ & $-2.684 \pm 0.199$ \\
    $0.146$ & $9.815  \pm 0.139$ & $-1.410 \pm 0.031$ & $-0.245 \pm 0.566$ & $-3.086 \pm 0.103$ & $-3.049 \pm 1.276$ \\
    $0.214$ & $9.622  \pm 0.022$ & $-1.349 \pm 0.016$ & $1.490  \pm 0.128$ & $-2.973 \pm 0.044$ & $-3.226 \pm 0.065$ \\
    $0.288$ & $9.871  \pm 0.209$ & $-1.339 \pm 0.036$ & $1.420  \pm 0.351$ & $-2.970 \pm 0.162$ & $-3.792 \pm 0.519$ \\
    $0.371$ & $9.810  \pm 0.116$ & $-1.272 \pm 0.033$ & $1.363  \pm 0.327$ & $-2.696 \pm 0.092$ & $-3.772 \pm 0.417$ \\
    $0.464$ & $10.288 \pm 0.102$ & $-1.404 \pm 0.045$ & $0.933  \pm 0.953$ & $-3.154 \pm 0.115$ & $-4.717 \pm 1.180$ \\
    $0.569$ & $10.208 \pm 0.087$ & $-1.287 \pm 0.036$ & $1.311  \pm 0.690$ & $-3.004 \pm 0.081$ & $-4.614 \pm 0.979$ \\
    $0.690$ & $10.224 \pm 0.111$ & $-1.202 \pm 0.081$ & $0.067  \pm 0.996$ & $-2.722 \pm 0.123$ & $-5.217 \pm 1.630$ \\
    $0.832$ & $10.290 \pm 0.112$ & $-1.188 \pm 0.088$ & $0.159  \pm 1.038$ & $-2.779 \pm 0.138$ & $-5.252 \pm 1.582$ \\
    $1.001$ & $10.558 \pm 0.126$ & $-1.262 \pm 0.053$ & $0.662  \pm 1.064$ & $-3.016 \pm 0.108$ & $-4.640 \pm 1.464$ \\
    $1.208$ & $10.617 \pm 0.155$ & $-1.179 \pm 0.060$ & $0.770  \pm 1.030$ & $-3.016 \pm 0.127$ & $-4.634 \pm 1.616$ \\
    $1.470$ & $10.639 \pm 0.149$ & $-1.146 \pm 0.057$ & $0.729  \pm 1.022$ & $-3.101 \pm 0.111$ & $-4.698 \pm 1.494$ \\
    $1.818$ & $10.725 \pm 0.063$ & $-1.218 \pm 0.042$ & $0.414  \pm 1.011$ & $-3.274 \pm 0.081$ & $-5.418 \pm 1.482$ \\
    $2.311$ & $10.990 \pm 0.183$ & $-1.473 \pm 0.088$ & $0.757  \pm 0.983$ & $-3.837 \pm 0.233$ & $-5.338 \pm 1.266$ \\
    $3.090$ & $10.728 \pm 0.069$ & $-1.169 \pm 0.075$ & $0.094  \pm 1.021$ & $-3.454 \pm 0.099$ & $-5.847 \pm 1.400$ \\
    \hline
    \end{tabular}
    \caption{Double Schechter function fit parameters and $1\sigma$ uncertainties for the GMFs. These fitted functions should be multiplied by the LSS correction factor get the same CGMH as in Figure~\ref{fig:rho} and Table~\ref{tab:rho_baryon}.}
    \label{tab:gmf_par}
\end{table*}

\begin{table*}
    \centering
    \begin{tabular}{ccccc}
    $\left< z \right>$ & $\mathrm{\log_{10}(\rho_{\star} \, [M_{\odot} \, Mpc^{-3}]})$ & $\mathrm{\log_{10}(\rho_{dust}\, [M_{\odot} \, Mpc^{-3}]})$ & $\mathrm{\log_{10}(\rho_{neutral} \, [M_{\odot} \, Mpc^{-3}]})$ & LSS correction factor [dex]\\
    \hline
    $0.027$ & $8.470 \pm 0.052$ & $5.189 \pm 0.055$ & $7.074 \pm 0.052$ & $0.141  \pm 0.043$ \\
    $0.084$ & $8.455 \pm 0.044$ & $5.192 \pm 0.046$ & $7.108 \pm 0.047$ & $0.024  \pm 0.037$ \\
    $0.146$ & $8.438 \pm 0.045$ & $5.197 \pm 0.045$ & $7.111 \pm 0.049$ & $-0.008 \pm 0.037$ \\
    $0.214$ & $8.421 \pm 0.041$ & $5.277 \pm 0.040$ & $7.153 \pm 0.043$ & $0.015  \pm 0.033$ \\
    $0.288$ & $8.402 \pm 0.043$ & $5.338 \pm 0.044$ & $7.203 \pm 0.040$ & $0.037  \pm 0.034$ \\
    $0.371$ & $8.380 \pm 0.040$ & $5.384 \pm 0.048$ & $7.147 \pm 0.043$ & $-0.153 \pm 0.032$ \\
    $0.464$ & $8.357 \pm 0.042$ & $5.429 \pm 0.061$ & $7.370 \pm 0.045$ & $0.041  \pm 0.033$ \\
    $0.569$ & $8.329 \pm 0.041$ & $5.462 \pm 0.046$ & $7.446 \pm 0.046$ & $0.100  \pm 0.033$ \\
    $0.690$ & $8.298 \pm 0.056$ & $5.411 \pm 0.070$ & $7.461 \pm 0.057$ & $-0.121 \pm 0.044$ \\
    $0.832$ & $8.259 \pm 0.055$ & $5.369 \pm 0.059$ & $7.482 \pm 0.054$ & $-0.115 \pm 0.044$ \\
    $1.001$ & $8.210 \pm 0.053$ & $5.453 \pm 0.069$ & $7.548 \pm 0.052$ & $-0.107 \pm 0.044$ \\
    $1.208$ & $8.146 \pm 0.057$ & $5.562 \pm 0.067$ & $7.631 \pm 0.057$ & $-0.045 \pm 0.048$ \\
    $1.470$ & $8.055 \pm 0.052$ & $5.431 \pm 0.063$ & $7.573 \pm 0.055$ & $-0.028 \pm 0.046$ \\
    $1.818$ & $7.919 \pm 0.051$ & $5.502 \pm 0.056$ & $7.529 \pm 0.052$ & $-0.008 \pm 0.046$ \\
    $2.311$ & $7.699 \pm 0.056$ & $5.235 \pm 0.064$ & $7.453 \pm 0.060$ & $0.037  \pm 0.051$ \\
    $3.090$ & $7.394 \pm 0.136$ & $5.345 \pm 0.138$ & $7.407 \pm 0.137$ & $0.064  \pm 0.134$ \\

    \hline
    \end{tabular}
    \caption{Measurements and $1\sigma$ uncertainties for the CSMH, CDMH and CGMH as in Figure~\ref{fig:rho}. Each of these columns and the uncertainties factor in the LSS corrections and their associated uncertainties, which are themselves presented in the rightmost column.}
    \label{tab:rho_baryon}
\end{table*}

\bsp	
\label{lastpage}
\end{document}